# Inelastic dilatancy as a mechanism for coseismic fluid depressurization of a shallow fault zone

Ruei-Jiun Hung[1,2], Matthew Weingarten[1], Shuo Ma[1], Steven M. Day[1]

1 Department of Earth and Environmental Sciences, San Diego State University, California, USA.

2. Scripps Institution of Oceanography, University of California, San Diego, California, USA.

1 Inelastic dilation causes amplified near-fault pore pressure declines, explaining large drawdowns better than poroelastic models.

2. A dynamic rupture model with inelastic dilatancy generates enhanced fault zone permeability distributed asymmetrically across the fault.

3. Coseismic fault zone depressurization draws groundwater inflow back into the fault zone during the postseismic phase.

## Abstract

Hydrologic observations and experimental studies indicate that inelastic dilation from coseismic fault damage can cause substantial pore pressure reduction, yet most near-fault hydromechanical models ignore such inelastic effects. Here, we present a 3-D groundwater flow model incorporating the effects of inelastic dilation based on an earthquake dynamic rupture model with inelastic off-fault deformation, both on pore pressure and permeability enhancement. Our results show that inelastic dilation causes mostly notable depressurization within ~1 km off the fault at shallow depths (< 3 km). We found agreement between our model predictions and recent field observations, namely that both sides of the fault can experience large-magnitude (~tens of meters) water level drawdowns. For comparison, simulations considering only elastic strain produced smaller water level changes (~several meters) and contrasting signs of water level change on either side of the fault. The models show that inelastic dilation is a mechanism for coseismic fault depressurization at shallow depths. While the inelastic dilation is a localized phenomenon which is most pronounced in the fault zone, the pressure gradients produced in the coseismic phase have a broader effect, increasing fluid migration back into the fault zone in the postseismic phase. We suggest field hydrologic measurements in the very-near-field (< 1 km) of active faults could capture damage-related pore pressure signals produced by inelastic dilation, helping improve our understanding of fault mechanics and groundwater management near active faults.

**Plain Language Summary**

Groundwater level data collected in the ruptured fault zone after an earthquake show significant water level drop followed by enhanced groundwater flow recharge. This can be understood as the opening of fractures that are well connected (fault-zone "damage"), allowing groundwater to fill the fracture voids and migrate efficiently. However, groundwater flow models addressing the fault-fluid problems often do not include the earthquake-induced damage in computing the groundwater response. Instead, these models consider the elastic deformation as the primary cause to the water level changes. In this study, we calculate groundwater response to a computer-simulated fault rupture that includes both elastic deformation and fault-zone damage. Our results indicate that (1) shallow damage can attract groundwater to flow into the fault zone as a result of fracture space creation, (2) groundwater level reduction is larger, and (3) the fluid migration is enhanced because new fluid pathways open in the fault zone. These modeled results cannot be captured by the elastic deformation model itself. Our model probably provides better understanding of fault-fluid interactions in response to earthquakes.

**Introduction**

Earthquake-induced hydrologic changes provide useful information to infer crustal pore pressure response, poroelastic parameters, hydrologic properties, as well as their temporal evolution. It has been recognized that, during an earthquake, fault rupture and seismic shaking can cause rock deformation or even damage, resulting in noticeable hydrologic response of groundwater level, geysers, and stream flow in both near- and far-fields (Wang, & Manga, 2010; see also Wang & Manga, 2021 for review). A commonly observed hydrologic phenomenon associated with earthquakes in the near-field (distance < 1 fault length, Wang & Manga, 2010) is that the pore pressure or water levels may experience a step-like change resulting from the sudden deformation associated with the fault slip. According to the theory of poroelasticity, the sign of the water level change, either increase or decrease, is determined by the sign of the coseismic volumetric strain. That is, dilational strain opens the pore space for a decreasing fluid pressure, whereas the compressional strain reduces the pore space thus increasing the fluid pressure (Ge & Stover, 2001; Muir-Wood & King, 1993; Roeloffs, 1996; Wang, 2000). Based upon this concept, numerical studies focused on earthquake-fluid interactions have reated the poroelastic response as the primary (and often sole) mechanism for coseismic water level change, without considering other effects such as permeability change and inelastic dilatation (e.g., Jónsson et al., 2003; Miao et al. 2021; Zhou & Burbey, 2014).

While a handful of observational studies supported the applicability of using poroelastic theory for the coseismic pore pressure prediction in near-field (e.g., Jónsson et al., 2003; Quilty & Roeloffs, 1997; Shi et al., 2013), many near-fault data have shown that the coseismic groundwater level (or pore pressure) changes often mismatch the prediction suggested by this theory (Lai et al., 2004; Hosono et al., 2019; Hung et al., 2024; Koizumi et al. 1996; Rojstaczer, 1995; Rutter et al., 2016; Wang et al., 2004b). For instance, the hydrologic observations during the 2016 $M_w$7.0 Kumamoto, Japan earthquake revealed that the groundwater level reductions occurred nearly everywhere in the ruptured fault system and are not consistent with the coseismic pore pressure changes predicted from poroelastic theory (Hosono et al., 2019). Similarly, near-fault observations for the 2018 $M_w$6.3 Hualien, Taiwan earthquake, showed coseismic water level drawdowns on both sides of the ruptured strike-slip fault, and do not agree with the poroelastic model (Hung et al. 2024). Consequently, other effects such as rock damage and hydrologic property changes may be required in the model to better explain these coseismic and postseismic groundwater responses.

In general, fault-zone fractures may exhibit preferential alignment and aperture opening/closure at certain orientations, resulting in an anisotropic permeability structure (Evans et al., 1997; Farrell et al., 2014; Faulkner & Armitage, 2013). Studies show that a typical anisotropic fault zone has its highest and lowest permeabilities parallel and normal to the fault interface, respectively (Evans et al., 1997; Farrell et al., 2014; Gudmundsson, 2000). A ruptured fault zone can act as a groundwater conduit due to the anisotropic permeability enhancement, facilitating fluid migration and circulation in these zones of enhancement (Gudmundsson, 2000; Morton et al., 2012; Sibson, 1990; 1994). During an earthquake rupture, fractures created by inelastic shearing processes can also lead to aperture opening (i.e., fracture dilation) as the colliding fracture roughness pulls against each other (Barton et al., 1995; Brown, 1987). Consequently, shear-induced inelastic dilation (fracture space creation) may significantly reduce the fault-zone pore pressure, particularly at shallow depths (Brantut, 2020; Scholz et al., 1973). Both of the mechanisms (i.e., pore pressure drop and permeability enhancement) serve as drivers for the enhanced fluid flows and are capable of modifying the postseismic stress state as well as aftershock characteristics (e.g., Cox, 2010; Nur & Booker 1972; Sibson, 1990; 1994).

While hydrologic data related to coseismic fault-zone response are somewhat limited, a few datasets successfully captured such a response and offer insights on the fluid-fault interactions during and after an earthquake. The groundwater level data associated with the 2018 Hualien, Taiwan earthquake provide a rare opportunity to study the coseismic fault-zone hydrologic response, because the groundwater wells were very close to the surface rupture of the strike-slip Milun fault (180-250 m, Fig. 1). The data provide two first-order observations: (1) groundwater level drawdown occurred on both sides of the fault and (2) the magnitude of water level drawdown was large (10-15 m) at two very-near-fault stations. Hung et al. (2024) argued that such phenomena cannot be well explained using the poroelastic theory. Instead, they modeled the postseismic groundwater flow using a step-like change in the hydrologic properties and prescribed a pressure sink in the fault zone to mimic the fault damage effects. Modeled water levels under this parameterization better matched observations than an elastic model alone.

Coseismic groundwater level/pore pressure drop in the fault zone is linked to the coseismic rock damage, which can be simulated using dynamic earthquake rupture models with off-fault yielding. Ma (2008) and Ma & Andrews (2010) showed that widespread inelastic strain (damage) accumulation in the strike-slip fault surroundings at near-surface depth, where the dynamic stress excited from seismic wave and rupture front permits off-fault yielding on both sides of the fault under the low confining stress conditions. Towards seismogenic depth, the increasing confining stress narrows down the damage zone width, forming a pattern analogous to the 'flower-structure' commonly observed in the field (i.e., Woodcock & Fischer, 1986). Such a finding may explain the extensive groundwater level drawdown observed during the Hualien earthquake, as the pore pressure reduction on both sides of the fault is likely induced by the inelastic strain and dilation at the shallow depths.

In this study, we further explore the effect of inelastic strain (damage) and the hydrologic response of a ruptured fault system in a 3-D model. While several modeling studies have also addressed the importance of inelastic dilation on the pore fluid response, they are either restricted to 2-D (Cappa et al., 2009; Matthäi & Fischer, 1996; Viesca et al., 2008; Zhang & Sanderson, 1996; Zhu et al., 2020) or prescribe pre-determined hydraulic parameters that are homogenously distributed in the fault zone (Matthäi & Fischer, 1996; Zhang et al., 2008; Zhu et al., 2020). Moreover, most of these models ignored contributions from seismic wave-induced dynamic stresses. The generation of coseismic damage may result from both fault rupture and seismic waves, leading to heterogenous damage distributions both inside and outside the fault zone (e.g., Ma, 2008; Ma & Andrews, 2010). Here, we use earthquake dynamic rupture simulation, which models the combined effect of fault rupture and seismic wave radiation to generate off-fault inelastic strain. We follow Ma (2008) and Ma & Andrews (2010) and investigate the impact of coseismic inelastic strain and pore pressure change on fault-zone groundwater flow. Unlike kinematic models, the dynamic rupture model calculates spontaneous fault rupture, coupled to the time-dependent, heterogenous distribution of stress, pore pressure, and inelastic strain. The outputs (i.e., undrained pore pressure and inelastic strain) from the 3D earthquake dynamic rupture model are translated into a 3-D groundwater flow model to further analyze pressure diffusion in the postseismic phase. Comparison of the pore pressure evolution from two models (inelastic versus elastic) provides insight into the role of fault-zone damage on coseismic and postseismic groundwater flow and pore pressure evolution, with implications for fault mechanics, hydrogeology, and groundwater resource management (e.g., Bense et al., 2013; Ingebritsen & Manga, 2019).

## 2 Methodology

### 2.1 Earthquake dynamic rupture model

Earthquake dynamic rupture models allow for the simulation of both coseismic fault damage and the undrained pore pressure response. Here, we construct our model and parametrization based on Ma (2008) and Ma & Andrews (2010) by prescribing an identical stress configuration and yielding criterion. However, we vary the model dimension and off-fault material properties to reflect a heterogenous and perhaps more realistic crustal environment. We designed a vertical fault with dimension $10 \times 30$ km$^2$ which is comparable with the Milun fault dimension, capable of generating a M6 earthquake (Fig. 2a). The maximum depth of the fault (10 km) is still above the brittle-ductile transition zone, in which the fluid may be overpressurized (Grawinkel & Stöckhert 1997; Sibson, 1983; Townend & Zoback, 2000). The fault length of 30 km roughly agrees with the width-length scaling relationship for continental strike-slip faults (Leonaed, 2014). The initial stress setting is identical to Ma (2008) and Ma & Andrews (2010), such that $\sigma_{xx} = \sigma_{yy} = \sigma_{zz} = (1 - \lambda)\rho g z$, where $\lambda$ is the pore pressure ratio, $\rho$ is the rock density, $g$ is the gravitational acceleration, and $z$ is depth; the shear stress $\sigma_{xy} = \sigma_{xx}$ and $\sigma_{xz} = \sigma_{yz} = 0$ form a high-angle maximum principal stress ($\sigma_{Max}$) orientation to the strike-slip fault (i.e., $\Psi$=45°; Fig. 2a&b). Such a high angle of $\sigma_{Max}$ permits the inelastic strain to dominate in the extensional quadrant at depth (Dunham et al., 2011; Ma, 2008; Viesca et al., 2008). The fault geometry and the initial stress settings are not specified for any particular earthquake but provide a representative earthquake rupture model for the hydrologic response assessment.

In our dynamic simulation the initial static friction $\mu_s$ = 0.6 and dynamic friction $\mu_d$ = 0.3 are applied on the fault plane. Tapering of $\mu_d$ is also applied at the fault boundaries with tapering width of ~2 km at each side so that the rupture can gradually slow down and avoid unrealistic strain concentration at fault edges (Fig. 2c). We set the initial shear-to-normal stress ratio $\mu_0$ to 0.4, so that the dimensionless factor controlling the sub- to supershear transition, S $= \frac{(\mu_s - \mu_0)}{(\mu_0 - \mu_d)}$, equals 2, appropriate for a sub-shear rupture (Andrews, 1976). We used a time-weakening friction law for the rupture nucleation and propagation (Andrews, 2004), during which the static friction linearly decreases to the dynamic friction over a characteristic time $t_c$ = 0.08 s, resulting in a varying characteristic slip distance $D_c$. The resulting increase of $D_c$ during rupture acceleration counteracts the natural scale compression associated with rupture propagation, ensuring that the breakdown zone is well resolved at all times on the discretized fault plane.

We adapted the 1-D velocity model from Boore & Joyner (1997) for the general crust but restricted the minimum shear-wave velocity = 1200 m/s to minimize numerical dispersion (Fig. S1a&b). The rock density is converted from the P-wave velocity based on the Nafe-Drake law (e.g., Brocher, 2005; Fig. S1c). The average density results in $\lambda$ ~0.383. In addition, we prescribed the internal friction $\mu = 0.7$ in most parts of the model, except for the shallow crust (< 2 km), where we linearly decreased the $\mu$ down to 0.6 to account for the weakened crustal rock observed in nature (e.g., Zoback & Townend, 2001, Fig. 3a). The use of minimum $\mu = 0.6$ is no greater than the static fault friction $\mu_s$.

We prescribe a low-velocity and low internal friction in the fault zone to represent the pre-existing fault structure before the dynamic rupture. It has been shown that fault zones are generally characterized as a low-velocity structure where enhanced fracture networks are observed as a long-existing phenomenon (e.g., Cochran et al., 2009). Generally, the total width of low-velocity fault zone ranges ~200-1500 m (Scholz, 2019), and we use 600 m (300 m on each side of the fault) in the model, consistent with some field observations (Rempe et al., 2018; Wu et al., 2019). Within the pre-existing fault zone, the P- and S-wave velocities are decreased exponentially based on the formula: $\left(V_{P,S} - V_{P,S}^f\right) exp\left(-\frac{|d|}{\zeta}\right)$, where $V_{P,S}$ and $V_{P,S}^f$ are the P- or S-wave velocity in the background and at the fault center, respectively. $d \leq 300$ m is the distance measured from the fault center (x = 0 km) within the damage zone, and $\zeta = 80$ m is a parameter controlling the changing rate of the seismic wave speeds in the fault zone. In our model, we assumed the $V_{p,s}^f$ is 30% lower than the background values (e.g., Cocco & Rice, 2002), and the width of the initial low velocity zone is constant (= $d$) over the entire fault along the depth. The setup of constant fault-zone width in the model is similar to other studies (e.g., Cappa, 2009; Zhu et al., 2020) which represents an idealized and simplified fault zone structure. We note that, however, real fault zones can also exhibit varying widths along the depth (i.e., flower structure). We lowered the mesh size to 60 m within 1 km of the fault plane to also minimize the numerical dispersion in the low-velocity zone. The internal friction $\mu$ is prescribed in a similar manner, in which we lowered the $\mu$ inside the fault damage zone and prescribed a minimum value of 0.6 (e.g., Scholz, 2000) at the fault zone center followed by the same exponential change.

The inelastic strain occurs when the invariant of deviatoric stress $\tau$ (presented as the square root of the second invariant, Eq.1) exceeds the Drucker-Prager yielding stress $\tau^y$ (i.e., $\tau \geq \tau^y$, Drucker & Prager, 1952), where $\tau$ and $\tau^y$ are expressed as:

$$\tau = \sqrt{0.5 s_{ij} s_{ij}},$$
$$\tau^y = -(\frac{\sigma_{kk}}{3} + P)sin\phi + c\,cos\phi, \qquad (1)$$

where $s_{ij}$ is the deviatoric stress, $\frac{\sigma_{kk}}{3}$ is the mean stress, $P$ is the pore pressure that can changes during the rupture. $\phi = tan^{-1}\mu$, and $c$ is the cohesion. We adapted the parameter of closeness-to-failure $CF$ to determine the likelihood of inelastic strain occurrence and magnitude (Viesca et al., 2008): $CF = \frac{\tau}{\tau^y}|_{t=0}$. As shown in Eq. 1, the prescribed $\phi$ and $c$ result in a range of $CF$ between 0 and 1, with higher value indicating the material is more prone to failure. We assume a cohesionless material (c = 0) at the depth shallower than 7 km in the model and thus $CF$ is controlled by $\phi$ only (herein, inelastic model), leading to a maximum initial $CF$ ~0.77 (Fig. 3b). However, we gradually increase cohesion at the depth greater than 7 km to avoid potentially artificial inelastic strain at the deep fault zone. Such a prescription does not affect the overall fault-zone damage behavior at shallow depth. A complementary model where no yielding is allowed (herein, elastic model) is also presented for comparison. In the elastic model we prescribed the identical velocity structure but used a large cohesion such that $CF = 0$ (Fig. 3c).

Once the yielding occurs, the inelastic shear strain $\gamma^p$ starts to accumulate at every time step. Below we follow the formulation derived by Hirakawa & Ma (2016) to obtain the inelastic strain increment and the subsequent pore pressure change. The inelastic shear strain increment is expressed as:

$$d\gamma^p = \frac{\tau - \tau^y}{h + G + K_u \beta sin\phi (1-B)^2 + \frac{KB\beta sin\phi}{\alpha}}, \qquad (2)$$

where $h$ is the hardening parameter. We take 20% of the maximum shear modulus in the velocity model ($G\sim33$ GPa), resulting in $h\sim6.6$ GPa (e.g., Hirakawa & Ma, 2016). $G$ is the shear modulus, $K_u$ and $K$ are the undrained and drained bulk modulus, respectively. $\beta = 0.1$ is the dilatancy defined as the ratio of inelastic volumetric-to-shear strains. The $\beta$ value here corresponds to soft rocks/loose materials dilation (e.g., Zhao & Cai, 2010). As shown in Section 4.5, we also discuss different $\beta$ (0.01-0.4) for the inelastic strain and pore pressure changes. The Skempton's coefficient $B$ is assumed to be 0.6 and the Biot's coefficient $\alpha$ is 0.5, both of which are suitable for the consolidated crustal materials (i.e., Wang, 2000). In Table S1 we list the parameters that were used in the model. Accordingly, the inelastic volumetric strain increment associated with the inelastic shear strain is expressed as:

$$d\varepsilon_{kk}^p = (1 - B)\,\beta d\gamma^p, \qquad (3)$$

which affects the undrained pore pressure change by the following equation:

$$dP = -B\frac{\sigma_{kk}}{3} + BK_u d\varepsilon_{kk}^p - \frac{KB\,\beta d\gamma^p}{\alpha}. \qquad (4)$$

Eq.4 denotes the pore pressure change responding to both elastic and inelastic deformations. Without the inelastic strains increment $d\varepsilon_{kk}^p$ and $d\gamma^p$, the pore pressure change $dP$ derives only from the first term which is the poroelastic response.

## 2.2 Coseismic permeability and hydrologic model setting

The final undrained pore pressure change $\Delta P$, which is the integrated $dP$ over rupture duration, and the volumetric strain $\varepsilon_{kk}^p$, are crucial for our postseismic groundwater flow modeling, because (1) $\Delta P$ may be used to infer the initial pressure head change due to the poroelastic deformation and rock dilation, and (2) $\varepsilon_{kk}^p$ may provide indications on the fracture volume and permeability change, as demonstrated by previous studies (Brace, 1966; 1978; Cappa, 2009; Johri et al, 2014; Peach & Spicer, 1996). Here, we referred to the study by Peach & Spicer (1996), in which they derived volumetric strain and fracture permeability relationship based on experiments and percolation theory:

$$k = \frac{\theta(\varepsilon_{kk}^p)^3 \pi^2 \langle r \rangle^2}{480 \Gamma}, \qquad (5)$$

where $\theta$ is hydraulic aperture shape and drag factor, describing the deviation of flow from the channeled Poiseuille flow, $\langle r \rangle$ is the mean radius of penny-like cracks, and $\Gamma$ is a volumetric shape factor defined as a ratio of tapered crack volume to an ideal penny-shaped crack volume. We follow that model, with $\theta = 0.5$ and $\Gamma = 2/3$. Eq.5 is valid when we assume the $\varepsilon_{kk}^p$ only causes the aperture opening without generating more fractures under the fluid-saturated condition. This equation estimates permeability evolution within the percolation regime where small $\varepsilon_{kk}^p$ increase can effectively enhance permeability. We assumed the $\langle r \rangle$ is 10 m, which is a size corresponding to macro-fractures (Chernyshev & Dearman, 1991) and may be reasonable for fault zones (e.g.,

Dor et al., 2006). We also provide a sensitivity test against different fracture dimensions in Section 4.5.

The coseismic $\Delta P$ and permeability changes from the earthquake dynamic rupture model are translated into a groundwater flow model. We simulated the postseismic fluid flow using MODFLOW 6, which solves the pore pressure diffusion equation based on Darcy's law in 3-D (Langevin et al., 2017):

$$\frac{\partial}{\partial x}\left(K_x^h \frac{\partial H}{\partial x}\right) + \frac{\partial}{\partial y}\left(K_y^h \frac{\partial H}{\partial y}\right) + \frac{\partial}{\partial z}\left(K_z^h \frac{\partial H}{\partial z}\right) = S_s \frac{\partial H}{\partial t}, \qquad (6)$$

where $K_x^h$, $K_y^h$, and $K_z^h$ are hydraulic conductivity along the x, y, and z directions, respectively. The conversion between permeability $k$ and conductivity $K^h$ is based on the relation: $K^h = \frac{k\rho_w g}{\xi}$, where $\xi$ is the fluid viscosity, $\rho_w$ is the water density, and $g$ is the gravitational acceleration (Freeze & Cherry, 1979). $S_s$ is the specific storage which we treated as a constant $=10^{-7}$ m$^{-1}$, and $H = \Delta P / \rho_w g$ is the pressure head change.

We created a 3-D model consisting of 1,546,704 unstructured grid elements (Fig. 4a) and prescribed the smallest grid size of 60 m around the fault (< 960 m) to preserve the same model resolution (Fig. 4b). The horizontal dimension of the model is approximately one fault length measured from the fault edge, and the vertical dimension is roughly the same as fault dip dimension (~10 km). The boundary condition is set to be $\Delta P = 0$, meaning no flow is allowed to enter/exit the model. All the grid cells are assumed saturated throughout the diffusion process and will never go dry. To account for the heterogenous distribution of permeability in the crust, we prescribed our initial permeability structure as a depth-dependent distribution (Fig. 4a). The initial 1-D permeability is based on Rice (1992): $k_h = k_s exp \left( \frac{-\sigma_n}{\sigma^*} \right)$, where $k_h$ is the background horizontal permeability ($k_h = k_x = k_y$), $\sigma_n$ is the normal stress on the fault, $\sigma^* = 30\ MPa$, and $k_s \sim 10^{-14}$ m$^2$ (e.g., Zhu et al., 2020). The background permeability structure from surface to depth in the model is important to the fluid flow process. While several studies prescribed their fluid flow model assuming an impermeable crust (e.g., Matthäi & Fischer, 1996; Zhu et al. 2020), others suggest that crustal permeability can be as large as $10^{-14}$ m$^2$ at the near-surface depths (Shmonov et al., 2003). Varying the background permeability can result in significantly different water level responses and therefore we tested three orders of magnitude in background permeability uncertainty from near-surface to 10 km depth: (1) $10^{-14}$ - $10^{-17}$ m$^2$, (2) $10^{-15}$ - $10^{-18}$ m$^2$, and (3) $10^{-16}$ - $10^{-19}$ m$^2$ (see additional details in Figure S3 & Table S2).

In addition, the layering of aquifers/aquitards in the crust forms an anisotropic permeability structure (e.g., Manning & Ingebritsen, 1999), also affecting the postseismic groundwater flow. Here, we set the vertical permeability $k_z$ uniformly 100 times smaller than the horizontal permeability $k_h$ (i.e., $k_z/k_h = 0.01$) in the model. Many studies suggest a ratio of $k_z/k_h \sim 0.1$ (Manning & Ingebritsen, 1999), which we tested with sensitivity analysis, but stronger background anisotropy generated more reasonable groundwater flow responses in the postseismic period. Such a setting may be expected when an aquifer/aquitard contrast is present in heterogeneous near-surface aquifers (see Section 4.3 & S1).

Fault damage-induced permeability enhancement and anisotropy may play an important role in the preferential groundwater flows (e.g., Gudmundsson, 2000). When there is inelastic strain taking place within a given model cell element, we added the calculated permeability value using Eq (5) to the background permeability to represent damage-induced permeability for both $k_h$ and $k_z$. Lastly, we included a fault-zone anisotropy factor in the inelastic model, where we gradually lowered the fault-crossing permeability $k_x$ by a factor of $10^{-4}$ (Evans et al.,1997) with respect to $k_y$ in order to mimic the fault-aligned vertical fractures at the fault center (Fig. S2). On the other hand, we do not specify such a fault-zone anisotropy before/after the earthquake in the elastic model, meaning that the elastic response without fracture generation similar to other poroelastic studies (e.g., Miao et al., 2021). Both coseismic $\Delta P$ and permeability changes are prescribed as step-like change at elapse time step = 0. The new permeability structure remains unchanged during the modeled postseismic period of two years with elapse time interval unit in days.

# 3 Results

## 3.1 Coseismic Plasticity and Pore Pressure Change

Here, we focus on the key outputs from the dynamic rupture simulation and compare the inelastic (with off-fault plasticity) and elastic (purely poroelastic) fault models. As discussed earlier, the primary motivation is to quantify how inelastic dilation in the shallow fault zone alters pore pressure relative to an elastic benchmark.

Fault rupture generates time-dependent coseismic slip, stress drop, inelastic strain, and pore pressure evolution in the fault zone (Fig. 5 and movies S1), and produces an earthquake with Mw~6.3 (Fig. 6). The modeled rupture spontaneously propagates along the fault plane, releasing stress over a time scale of a few seconds. By the end of this rupture, both the inelastic and elastic models achieve similar total seismic moments. However, consistent with previous studies (e.g.,

Ma, 2012; 2022; Roten et al., 2017; Taufiqurrahman et al., 2023), we observe a shallow slip deficit in the inelastic model (Fig. 6a). This deficit is partly because rock damage and yielding at shallow depths (~0–2 km) dissipate some mechanical energy and reduce slip accumulation there, whereas most slip remains concentrated at deeper levels with less pronounced fault-zone yielding. In addition, the inelastic strain energy causes a slower rupture velocity (Fig. S3). In dip-slip fault systems such as subduction zones, the off-fault yielding reduces near-trench slip while retaining similar slip distribution in the deep fault plane (i.e., shallow slip deficit). The resulting off-fault inelastic inelastic deformation can contribute to vertical uplift of the overriding plate, serving as an effective mechanism for generating significant seafloor deformation despite the reduction in fault slip (Ma, 2012).

We plot the coseismic inelastic strain on the elements adjacent to the fault interface to show the on-fault damage distribution (i.e., at x = -30 m, Fig. 7a&b). The inelastic strain distribution on the shallow fault plane extends almost along the entire fault strike with slightly higher strain magnitude in the fault extensional quadrant. Such an uneven distribution is clear at depth (> 2 km), where the inelastic strain accumulation is pervasive along the extensional quadrant compared to that in the compressional quadrant. This is because the mean stress in the extensional quadrant decreases, thus lowering the yielding stress, whereas the mean stress in the compressional quadrant increases the yielding stress (Eq.1). We note that the degree of damage-zone asymmetry is less pronounced in our inelastic model compared to those presented in Ma (2008) and Ma & Andrews (2010), likely due to the existence of an initial low velocity zone which allows trapped wave reverberations, and therefore more symmetric coseismic damage within the fault zone.

We found that a significant amount of inelastic strain is concentrated in the shallow fault zone (< 2 km), with maximum magnitudes of about $10^{-3}$ and $10^{-4}$ for the shear and volumetric strains, respectively. The plastic yielding branches outward near the surface and extends at least 1 km off-fault (Fig. 8a&b), leading to a 'flower structure' pattern (i.e., Woodcock & Fischer, 1986). Similar to Ma (2008) and Ma & Andrews (2010), our results show that the inelastic strain appears on both side of the fault zone due to strong dynamic stress perturbation, producing a distinct inelastic strain distribution compared to other studies using quasi-static fault slip models (i.e., Cappa, 2009; Zhang et al., 2008). The quasi-static models produced inelastic strain by slip-induced deformation, and the resulting dilation is only accumulated on one side of the strike-slip fault, whereas our model includes dynamic rupture and seismic wave perturbations, generating inelastic strain accumulation on both sides of the fault. We note that the effects of dynamic stress from seismic wave and rupture are important to the extensive off-fault damage, reflecting a more complete representation of an earthquake rupture process.

Following the approach described in Section 2.2, we link inelastic volumetric strain to permeability by assuming that fracture apertures and permeability increase in proportion to the dilational strain (e.g. Brace, 1978; Cappa, 2009). Fault-zone permeability in the inelastic model can reach values of $\sim 10^{-14}$ $m^2$ mostly in the shallow fault zone (< 2 km, Fig. 7c & Fig. 8c). Towards deeper parts of the fault, the increasing confining stress discourages rock yielding and limits permeability enhancements to lower values. These results confirm that shallow inelastic dilation is a key mechanism for raising fault-zone permeability by one or more orders of magnitude. These enhancements are also evident in the off-fault damage region, though the effect diminishes laterally away from the fault plane and with depth (Fig. 8c).

The inelastic volumetric strain generated by rock dilatancy leads to a notable pore pressure decline in the fault zone compared to the elastic fault model. In Fig. 9a&b, we plotted the pore pressure change in the elements adjacent to the fault interface (i.e., x = -30 m) for both models and the difference, inelastic minus elastic, in Fig. 9c. Although both models generated similar 'bulge-like' pressure responses at deeper levels (~4-8 km), the shallow fault (< 2 km) indicates that coseismic depressurization effectively turns the damaged fault zone into a 'pressure sink' (Fig. 9a&10a). In contrast, the purely elastic model shows smaller amplitude changes and even an opposite-sign (increased) pore-pressure region on one side of the fault. We found that the inelastic model produced a uniformly distributed pressure sink along the shallow fault zone and extended pore pressure reduction in the off-fault regions (Fig. 10a), whereas the pore pressure polarity is flipped in the elastic model determined by the poroelastic deformation (Fig. 9b) with much smaller pore pressure changing magnitudes at the near-surface depth (Fig. 10b).

## 3.2 Groundwater flow induced by fault damage

The pressure gradients and permeability changes produced in the inelastic fault zone have a direct effect on both coseismic and postseismic groundwater flow. As outlined in Section 2.2, we input our modeled permeability fields into a 3-D groundwater flow model to track fluid migration over the postseismic timescales of 2 years.

In the inelastic model, the significant pore pressure drop produced by fault damage draws fluid into the shallow fault zone from two source regions: (1) the deeper fault zone and (2) aquifers in the shallow off-fault region. The dominant driver of fluid flow into the shallow fault zone is the pressure gradient created during the coseismic phase. However, enhanced vertical permeability (anisotropy) near the fault center also increases the vertical component of flow, helping fluid migrate rapidly into the depressurized zone. Fig. 11 compares the instantaneous flow for the two models in the coseismic period (elapse time step = 0 in the groundwater flow simulation), where the pore pressure just starts diffusing. The flow vectors are derived based on Darcy's law $q_i = -K_i \frac{\partial P}{\partial L_i}$ with $i$ denoting one of the x, y, z directions. In near-fault regions (x = ±1.2 km, y = ±15 km, z < 3 km), the average Darcy flux is found $2.4 \times 10^{-9}$ m/s, about 1.7 times greater than in the elastic model ($1.4 \times 10^{-9}$ m/s) due to the enhanced hydraulic conductivity and gradient created in the inelastic model. Note that the flow vectors are pointing horizontally because of the small $k_z/k_h$ ratio in the background permeability model.

We simulated groundwater levels at station arrays that are normal to the fault strike located at different fault-strike locations. Specifically, we focused on the water levels at y = 4 km (near the epicenter), 10 km, and 14 km (close to the fault edge) for changes in hydraulic head at depth of 150 m in the model. At the near-fault stations (x = ±30 m), the inelastic model produces a wide range of water level declines roughly between 20–90 m (Fig. 12a-d). This amplitude is not replicated in the elastic model, where changes are only a few meters (Fig. 12g-j). Both models show much smaller changes moving towards the fault edges at y = 14 km, where reduced slip and minimal inelastic strain occur. (Fig. 12e-f, Fig.12k-l). Notably, water level declines appear on both sides of the ruptured fault in the inelastic model – mirroring field observations in some near-fault wells (Hosono et al. 2019; Hung et al., 2024), whereas the elastic model often produces a contrast in sign of water level change across the fault plane (e.g., Miao et al., 2021). Finally, the inelastic model reveals localized zones of over-recharge or discharge caused by intricate fluid diffusion pathways and vertical flow from deeper crust (e.g., Fig. 12c). Such behavior qualitatively aligns with a range of field observations (e.g., Sibson, 1994; Hosono & Masaki, 2020; Hosono et al., 2020; see Section 4.3), reinforcing the idea that coseismic fault zone depressurization can draw significant fluid from both adjacent aquifers and deeper crustal reservoirs.

## 4 Discussion

### 4.1 Inelastic strain produced by earthquake dynamic rupture

The interaction between on-fault and off-fault dynamics results in different inelastic strain patterns. Either changing the earthquake source parameters or off-fault materials can lead to different results. In this study, we focus on the analysis regarding off-fault material properties, comparing both the pore pressure and hydrologic response with/without off-fault inelastic strain under the same earthquake source parameters. However, we note that the initial stress setup can be an important factor determining the off-fault inelastic strain pattern as well (e.g., Ma & Andrews, 2010; Viesca et al., 2008; Xu et al., 2012; Dunham et al., 2011; Gabriel et al., 2013). In particular, the magnitude and spatial distribution of the inelastic strain can vary depending on the maximum principal stress orientation $\Psi$, defined as an acute angle between the maximum principal stress and the fault strike (Fig. 2a). Viesca et al. (2008) and Dunham et al. (2011) showed that inelastic strain occurs on the fault extensional quadrant if $\Psi$ is greater than 30°. However, inelastic strain can also accumulate in the compressional quadrant if $\Psi < 20°$, in which the effect of deviatoric stress exceeds the mean stress on yielding. As demonstrated by Ma (2022), small $\Psi$ angle (15°) can generate extensive inelastic strain in the extensional side of the fault shallow depths. This indicates that the rock yielding on both sides of the fault may still occur at shallow depth even if a different $\Psi$ is given, because the rock yielding is caused by the dynamic stress at the low confining stress condition, leading to depressurized fault zone regardless of the principal stress configuration. Since we follow the parametrization from Ma (2008) and Ma & Andrews (2010), the off-fault yielding is predominant in the extensional quadrant at $\Psi = 45°$.

In addition, Xu et al. (2012) noted that crack-like fault rupture can result in extended, triangular-shaped regions of off-fault yielding, whereas pulse-like rupture creates narrower yielding zone with constant width. While models with different initial stress conditions will produce flow behaviors that differ from the current study, those models will be affected to a similar degree by inelastic dilation. Analysis of the different initial stress settings and the resultant fluid flow behaviors may be worth future exploration, as it is beyond the scope of this study.

## 4.2 Coseismic permeability from the inelastic model

Permeability enhancement associated with earthquake damage has been widely documented in various fault zones from several earthquakes (Lai et al., 2004; Hosono et al., 2020; Hung et al., 2024; Rojstaczer et al., 1995; Tadokoro et al., 2000; Tokunaga, 1999; Wang et al., 2004b). In principle, fluid migration through a damaged fault zone can be simulated using discrete fracture network models, where individual fracture geometries and apertures are explicitly specified (i.e., Caine & Forster, 1999). For large-scale seismogenic faults, however, such discrete modeling is computationally prohibitive. Our approach instead treats inelastic volumetric strain as indicative of bulk fracture aperture changes (Eq. 5), following previous work suggesting that small inelastic strains can translate into large permeability gains (Brace, 1978; Peach & Spicer, 1996). By using this relationship, we bypass the need to specify fracture density or connectivity explicitly, making simulation of crustal-scale fluid flow more tractable.

We find that fault zone permeability enhancement in the inelastic model reaches $\sim 10^{-14}$-$10^{-15}$ m$^2$ inside the damage zone and $\sim 10^{-15}$-$10^{-16}$ m$^2$ in the off-fault regions. These estimates generally agree with both in-situ and experimental studies (e.g., Aben et al., 2020; Wang et al., 2004a; Rojstaczer et al., 1995; Tadokoro et al., 2000). For example, Aben et al. (2020) measured permeability evolution of crystalline rocks subject to dynamic fracturing and showed that coseismic permeability can increase by several orders of magnitudes, reaching up to $\sim 10^{-13}$ m$^2$ in the damage zone. Seismicity migration and in-situ coring of the ruptured Nojima fault after the 1995 Kobe earthquake also exhibited an enhanced permeability of $\sim 10^{-14}$-$10^{-15}$ m$^2$ consistent with our modeled permeability range (Lockner et al., 2009; Tadokoro et al., 2000). In addition, the off-fault damage due to seismic wave perturbations was shown to enhance crustal permeability by $\sim$one order of magnitude after the 1989 Loma Prieta earthquake (e.g., Rojstaczer et al., 1995).

Our results further show strong spatial heterogeneity in fault zone permeability, especially in the dilatational side of the shallow rupture (Fig. 7c). Higher permeability enhancement is observed on this side of the rupture. Such an asymmetric permeability pattern is reflected by the uneven damage generated across the fault and aligns with field observations made in large strike-slip fault (Wibberley & Shimamoto, 2003).

## 4.3 Comparison of modeled groundwater levels and field observations

Changes in groundwater level, streamflow, and other hydrologic signals can provide valuable proxies for near-fault pore pressure variation (e.g., Hosono et al., 2019; Wang et al., 2004a; Wang & Manga, 2015). During the 2018 **M**6.3 Hualien earthquake, water levels dropped on both sides of the strike-slip Milun fault by over 10 m in some near-fault wells (Hung et al., 2024). Our inelastic model reproduces this pattern of 'universal drawdown' with magnitudes up to tens of meters (Fig. 12a-f), whereas the elastic model predicts much smaller changes and flipped signs across the fault (Fig. 12g-l). Moreover, our model results suggest that large water level declines are spatially limited to within a few hundred meters to ~1 km of the damage zone (Fig. 13). This localization arises from high inelastic strain in the shallow fault and is broadly consistent with laboratory experiments (Brantut, 2020; Grueschow et al., 2003). Farther away, the influence of damage diminishes, and the elastic (poroelastic) effects begin to play a more significant role in the pressure response at the broader scale (e.g., Albano et al., 2017; Jónsson et al., 2003; Shi et al., 2013).

Our model results generally show large coseismic water level drops (tens of meters) in the fault-zone region. While groundwater level drops have been reported for several earthquakes, they showed a wide range of drawdown magnitudes due to several factors: (1) fault damage style, (2) source-well distance, (3) measurement depth, and (4) site effects. In Table 1 we summarize some documented coseismic groundwater responses from the near-fault data, and a comparison to our model results. An extreme case was reported by Tokunaga (1999), documenting water level declines of ~70 m about 3 months after the Kobe, Japan, earthquake. The measurement was collected from an in-situ survey at the well near the ruptured fault. Although there is no continuous water level data in Tokunaga's study for comparison, our modeled water curves seem able to intersect a similar drawdown range at around 90 days (Fig 12a-d).

During the 2018 M6.3 Hualien earthquake, water levels dropped on both sides of the strike-slip Milun fault by over 10 m at near-fault stations Fuqian and Huagangshan, between 180 – 250 m from fault (Hung et al., 2024). Our inelastic model reproduces this pattern of drawdown on both sides of the fault in the immediate vicinity of the damage zone (within ~ 1 km), with magnitudes up to tens of meters (Fig. 13). This contrasts with the elastic model, which predicts much smaller changes and typically flipped signs on either side of the fault. However, we note that field observations further from the Milun fault, at Beipu and Tzuchi stations between ~2 and 2.5 km from the fault, also showed moderate drawdown (Fig. 1a). At these distances, our inelastic model predicts a transition back to the elastic response with pressure increase in the relevant quadrant, suggesting that the effect of inelastic dilation diminishes quickly past ~1 km from the fault (Fig. 13). While increasing dilatancy or crack dimension in the inelastic model may extend the drawdown range away from the fault (Fig. S10 & S11), the effect of inelastic dilation appears to be most important in the very near-field region ($< 1$ km). The very near-fault ($x = 30$ m) drawdowns ($> 80$ m) in our model likely represent an upper-bound estimate, since to our knowledge there are no field observations at such distances during coseismic rupture. In practice, rapid fracture healing or localized fault-core sealing could reduce the effective drawdowns observed over time in nature.

Ideally, our inelastic model would be calibrated to a site-specific simulation of the 2018 Hualien earthquake. However, we opted for a qualitative model comparison to ensure the results remain both robust and physically interpretable. This modeling choice was made to avoid overfitting as

(1) the rupture process and seismotectonics of the entire Milun fault rupture are complex, and (2) our hydrologic data are spatially sparse with four stations sampling only the shallow crust ( < 200 m) in the southern part of the fault region. Consequently, by successfully reproducing the signal polarity on both sides of the fault within 1 km, our generic model isolates inelastic dilatancy as the primary mechanism driving the very-near-fault hydraulic response. The discrepancy at distances of 2 – 2.5 km suggests the actual zone of off-fault damage or hydraulic complexity in the Hualien event may be wider than the simplified zone prescribed in our study.

The variation in drawdown curves (e.g., maximum amplitude, duration etc.) can be affected by several factors, including background permeability structure and the postseismic permeability healing. For the background permeability structure, we performed a series of sensitivity tests and found that a low permeability crust ($10^{-16}$-$10^{-19}$ m$^2$) slows down the postseismic recharge, resulting in prolonged recovery and enhanced water level drawdown. On the other hand, a permeable crust ($10^{-14}$-$10^{-17}$ m$^2$) accelerates pore pressure re-equilibrium, reducing both the drawdown amplitude and duration (see Supporting Information S1 & Fig. S4). In addition, an isotropic background permeability structure (i.e., $k_z/k_h$ = 1) promotes vertical fluid migration of deep pore pressure changes, causing substantial over-recharge on the compressional side of the fault and pore pressure reduction on the dilational sides of the fault. Stronger permeability anisotropy, $k_z/k_h$, impedes deep-fluid migration to the surface, which contribute to stronger postseismic water level changes in the shallow zone (Fig. S4).

Another factor that controls the postseismic fluid flow is whether the permeability healing is involved. With the consideration of permeability healing (or reduction) with time, we would expect that the magnitude of postseismic water level change, as well as the duration may be diminished (Matthäi & Fischer, 1996). Consequently, our models may be considered an upper-bound estimate on the longevity of both permeability change or large drawdown, because we do not incorporate the effects of fracture clogging. We reference hydrologic observations from the 2010 M7.0 Darfield, New Zealand earthquake as an example, for which substantial water level increase was observed in the fault-zone region and attributed to permeability reduction induced by rapid fracture clogging (Rutter et al., 2016). Simulation of permeability healing is challenging since it is highly sensitive to local geology site effects and poorly understood (e.g., Manga et al., 2012).

## 4.4 Fluid flow in damaged fault and implications

Despite the localization of the dilation and pore pressure drop in the fault zone, the pressure gradient and permeability enhancement can significantly impact the regional fluid flows. As shown in Fig. 11a, crustal fluid migrates toward the depressurized fault zone which acts as a pressure sink in the postseismic period. We note that such a fault depressurization can not only reduce the shallow aquifer pressure, but also likely drive deep fluid upwelling to shallow depth along the permeable damage zone if we assume the background $k_z/k_h = 1$. Faults drawing fluids upward from deeper levels has been suggested by previous studies (Brantut, 2020; Grueschow et al., 2003; Zhang & Sanderson, 1996). The mechanism resembles Sibson's (1994) "seismic suction pump", which has implications for fault stability/earthquake rupture arrest (Aben & Brantut, 2021; Liu & Duan, 2014; Sibson, 1989) and long-term mineral deposition in fault jogs (e.g., Cox, 1999).

While our study is focused on the shallow (< 3 km) pore fluid response shortly after the earthquake (< 2 years), it is also worth noting the hydrologic impacts on near-surface groundwater resources. Rupture-induced damage can breach aquitards, promoting rapid redistribution of fluid pressure and groundwater (Fleeger, 1999; Hosono et al., 2016; Liao et al., 2015; Rojstaczer et al., 1995; Wang et al., 2004b; Wang & Manga, 2015). For example, hydrologic analysis after the 2016 Kumamoto earthquake indicated that groundwater release from near-fault mountains quickly recharged the depressurized fault-zone aquifer, resulting in an elevated water table reaching above the preseismic levels (Hosono et al., 2019; 2020). On the other hand, enhanced flow paths can also drain local aquifers, exacerbating water shortages near the earthquake epicenter (i.e., Fleeger, 1999). In addition, groundwater contamination may also increase if damaged infrastructure or fracture pathways allow pollutants to migrate (e.g., Ingebritsen & Manga, 2019). As reported by Nakagawa et al. (2020), sewage pipes damaged by the fault rupture during the Kumamoto earthquake caused chemical leakage to the regional groundwater system and brings our awareness of future geohazard assessment regarding the groundwater supply in the fault damage area. The broader impacts of hydrologic changes produced by fault rupture may persist well beyond the 2-year postseismic window modelled in this study.

### 4.5 Sensitivity test of the dynamic parameters

The magnitudes of inelastic strain and permeability enhancement depend on material parameters. Such a discussion has been provided by Viesca et al. (2008), where they suggested the Skempton coefficient $B$ and dilatancy $\beta$ are important to the inelastic strain. Ma & Andrews (2010) also showed the effect of material cohesion $c$ to the inelastic strain generation. Here, we performed sensitivity tests by varying Skempton coefficient ($B$), cohesion ($c$) and dilatancy ($\beta$) to understand the expected range of pore pressure and permeability changes.

The Skempton coefficient $B$ describes the coupling between stress and pore pressure (e.g., Eq. 4). Fig. S5 shows the sensitivity results for $B$ = 0.60, 0.75, and 0.90. As the larger $B$ values indicate stronger pressure-stress coupling, it produces a greater pore pressure drop in the fault zone (Fig. S5). As shown in Viesca et al. (2008), the larger Skempton coefficient can promote and demote inelastic strain generation in the compressional and dilational quadrants, respectively, resulting in variations in the damage asymmetry. In our models, the presence of pre-existing low-velocity zone produces less pronounced damage asymmetry. While an enhanced pore pressure reduction is produced for the large Skempton coefficients, the postseismic maximum water level change is less amplified (Fig. S7), possibly due to the less pronounced permeability enhancement.

For the cohesion test (Fig. S7), we found that increasing $c$ = 1 MPa prevents the rock from substantial yielding, resulting in a pore pressure response nearly identical to the elastic model with limited permeability enhancement generated in the model. When cohesion is less than 0.1 MPa, the results of inelastic strain are similar to the cohesionless inelastic model, and the maximum water level change pattern becomes similar to the cohesionless inelastic model (Fig. S8).

We also performed sensitivity test for the dilatancy parameter $\beta$ (Fig. S9). We found that an increasing $\beta$ amplifies the depressurization while reducing the inelastic shear strain (e.g., Viesca et al., 2008). The inelastic volumetric strain shows similar patterns despite some local magnitude enhancements are observed. In addition, the higher $\beta$ value also generates broader extension of shallow volumetric strain and pressure drop at near-surface depth (Fig. S10). This results in a large water level change in the postseismic period, where the maximum water level drop can be as large as ~140 m within the fault zone ($\beta = 0.4$, Fig. S10a&b), and a widened area of water-level drop across the fault. Overall, the shape of maximum water level change curves is similar among each other (Fig. S10).

Lastly, we tested the effects of crack dimension (i.e., mean crack radius $\langle r \rangle$, Eq. 5) on the resultant groundwater flow. According to Chernyshev & Dearman (1991), typical fracture length falls between 0.01-10 m, whereas mega-scaled fractures or faults are in an order of 100 m or greater. Coseismic damage of fault zones often generate fractures with lengths between 1-100 m, and we therefore test the sensitivity of fracture size to the permeability and groundwater flows by setting $\langle r \rangle = 1$ m, 10 m, and 100 m. If we use $\langle r \rangle = 1$ m, Eq. 5 decreases the Darcy flow rate and produces narrower water drawdown range across the fault, whereas increasing $\langle r \rangle = 100$ m can produce widespread drawdown range (Fig. S11). However, increasing $\langle r \rangle = 100$ m does not necessarily induce larger water level drawdown magnitude in the fault zone, likely because of the faster fluid recharge which can quickly equilibrate the fault-zone low pressure. Indeed, we found that using $\langle r \rangle = 100$ m in Eq. 5 causes much faster Darcy flow of $7.98 \times 10^{-8}$ m/s in the fault zone region (x = $\pm 1.2$ km, y = $\pm 15$ km, z < 3 km) compared to the case of $\langle r \rangle = 10$ m ($2.4 \times 10^{-9}$ m/s) in the coseismic period. On the other hand, the smaller $\langle r \rangle$ of 1 m results in slower flow rate of $1.64 \times 10^{-9}$ m/s, whose value falls between the inelastic and elastic models.

## 4.6 Model limitations

While our earthquake dynamic rupture and groundwater models illustrate how near-fault damage drives large pore pressure and water level changes, we discuss some model limitations which inform our interpretations.

First, we do not include a fault core layer for the earthquake dynamic rupture model. The structural and mineral composition of the fault core can vary differently among different fault systems and results in complex coseismic responses such as thermal pressurization, inelastic dilation, and compaction (e.g., Hirakawa & Ma, 2016; Lockner & Byerlee, 1994; Proctor et al., 2020; Rice, 2006; Yao et al., 2023). While these processes involve substantial pore pressure change, they generally occur in the very localized area inside the fault core, and the pore pressure response mostly affects the seismic slip behavior rather than the regional groundwater system. Our groundwater model partially addresses this by imposing an anisotropic permeability barrier for fault-crossing fluids, effectively impeding fluid flow across the fault interface similar to the effect of a fault core (e.g., Gudmundsson, 2000; Faulkner & Armitage, 2013).

Second, while our earthquake dynamic rupture model considers the poroelastic response in the undrained condition, the subsequent groundwater flow model only solves the uncoupled pore pressure diffusion without re-coupling to elastic stress. It has been shown that pore pressure change can modify the elastic stress configuration and thus impact the aftershock activity (e.g., Albano et al., 2017), but the elastic stress modification only plays a tangential role on the pore fluid flows (e.g., Wang, 2000). Since our study is focused on the purely hydrologic response to the ruptured fault, the use of the uncoupled pore pressure diffusion equation (i.e., Darcy's law) is appropriate for our objectives.

Third, Eq. (5) assumes the permeability change is generated within the percolation regime, where small inelastic strain can lead to significant permeability change. When fractures are well connected (i.e., the connected regime), adding more inelastic strain will only cause slight permeability increases (e.g., Guéguen & Schubnel 2003; Zhu & Wong, 1999). According to experimental study results (e.g., Peach and Spicer, 1996; Zhu & Wong, 1999), the connected regime occurs when the volumetric strain reaches an order of $10^{-3}$, which is significantly higher than the inelastic volumetric strain of in our models ($\sim 10^{-5}$), consistent with Eq (5). We note that, however, permeability evolution during coseismic damage likely involves both percolation and connected processes in nature which set limits on earthquake-induced permeability. If the connected process initiates at a lower strain threshold, it is possible that our results indicate an upper-bound estimate of earthquake-induced permeability change.

In addition to permeability change, specific storage can also vary during coseismic fault damage. Specific storage is defined as $S_s = \rho_w g \left( C_m + n C_f \right)$, where $C_m$ is the material compressibility, $n$ is porosity, and $C_f$ is fluid compressibility. When fault-zone material is subject to damage, both $C_m$ and $n$ are expected to increase (Heap et al., 2010; Yang et al., 2021), leading to $S_s$ enhancement. We adopt the approach by Chen et al. (2023), where they evaluated elastic moduli change for different types of rock damage. We show that $S_s$ increase is negligibly small, thus unlikely to affect the postseismic fluid flow pattern (see Supporting Information S2 & Fig. S12 for detail). As a result, we focus on the permeability effects as the primary hydrologic property control to the fault-zone fluid flow.

Lastly, we do not model liquefaction in unconsolidated materials, nor do we allow for time-dependent permeability healing or clogging (e.g., Manga et al., 2012; Rutter et al., 2016). These contrasting phenomena can be understood as the consequence of temporal permeability change that is not incorporated in our simulation (see Sec 4.3). While they can significantly affect the postseismic fluid flow behaviors, they often contribute as a secondary effect induced by the fault damage and nonlinear soil response.

## 5 Conclusions

We simulated the hydrologic response to coseismic rupture of a strike-slip fault zone with inelastic off-fault deformation (damage) inferred from a dynamic earthquake rupture model. We compared these simulated responses to those of a dynamic rupture considering only elastic deformation. Distinct contrasts were found between the respective coseismic pore pressure distributions for the inelastic and elastic models, caused by fault depressurization and permeability enhancement. Here, we summarize the main findings:

1. The earthquake dynamic rupture model with rock yielding showed heterogenous permeability distributions both inside and outside the fault. Following the modeled inelastic strain pattern, damage-induced permeability enhancement reached values as high as $10^{-14}$ $m^2$ in the shallow (depth < 3 km) fault-zone with an asymmetric distribution across the strike-slip fault (e.g., Wibberley & Shimamoto, 2003).

2. Notable pore pressure declines, in excess of 0.1 MPa, can be generated on both sides of the shallow strike-slip fault (depth < 3 km) as a result of the inelastic dilation. This explains the data observed in some very near-fault groundwater stations and provides an alternative explanation for the water level polarity mismatch predicted by poroelastic models in near-fault regions (e.g., Hosono et al., 2019; Hung et al., 2024). The depressurized shallow fault zone attracts the surrounding groundwater flows into the fault damage zone at enhanced flow rates in the postseismic period.

3.   A hydrologic model incorporating fault damage effects (i.e., inelastic dilation) enables a larger groundwater drawdown compared to those induced by the poroelastic models and sensitivity analyses show comparable results with a wide range data observed in the field (e.g., Tokugawa, 1999; Hosono et al., 2019; Wang et al., 2004b; Hung et al., 2024). The enhanced water level reductions appear to be a localized phenomenon, forming a funnel-shaped pore pressure sink in the shallow, near-fault region.

While fault depressurization is a localized phenomenon, the enhanced pressure gradient and permeability permit more efficient fluid flow and aquifer communication in the crust, thus raising our concern for the threat to localized shallow groundwater contamination or depletion in the postseismic period. We suggest near-fault (distance < 1 km) hydrologic measurements may be important to infer detailed fault-fluid behavior in response to an earthquake rupture and may provide improved understanding of fault-zone architecture and postseismic groundwater flow behavior.

**Open Research**

The Fortran code for the earthquake dynamic rupture model is provided at Hung (2025). Groundwater flow model MODFLOW 6 is an open software provided by the US Geological Survey (Langevin et al., 2017). Figures and related processing are generated by Matlab_R2025.

## Acknowledgement

We thank editor Alexandre Schubnel and two anonymous reviewers for the constructive feedback. We also thank the associated editor for reviewing this study. The earthquake dynamic rupture modeling is supported by the Computational Science Data Center at the San Diego State University for the high-performance computing resources.

## Conflict of Interest Statement

The authors have no conflicts of interest to disclose.

## References

Aben, F. M., M.-L. Doan, & T. M. Mitchell (2020). Variation of hydraulic properties due to dynamic fracture damage: Implications for fault zones. *Journal of Geophysical Research*, **125**, e2019JB018919. https://doi.org/10.1029/2019JB018919

Aben, F. M., & N. Brantut (2021). Dilatancy stabilises shear failure in rock. *Earth and Planetary Science Letters*, **574**, 117174. https://doi.org/10.1016/j.epsl.2021.117174

Albano, M., S. Barba, G. Solaro, A. Pepe, C. Bignami, M. Moro, M. Saroli & S. Stramondo. (2017). Aftershocks, groundwater changes and postseismic ground displacements related to pore pressure gradients: Insights from the 2012 Emilia-Romagna earthquake. *Journal of Geophysical Research,* **122**(7), 5622-5638. https://doi.org/10.1002/2017JB014009

Andrews, D. J. (1976), Rupture velocity of plane strain shear cracks, *Journal of Geophysical Research*, **81**(32), 5679–5687. https://doi.org/10.1029/JB081i032p05679.

Andrews, D. J. (2004). Rupture models with dynamically determined breakdown displacement. Bulletin of the Seismological Society of America, **94**(3), 769–775. https://doi.org/10.1785/0120030142

Brown, S. R. (1987). Fluid flow through rock joints: The effect of surface roughness: Journal of Geophysical Research, **92**, p. 1337–1347. https://doi.org/10.1029/JB092iB02p01337

Barton, C. A., M. D. Zoback, & D. Moos (1995). Fluid flow along potentially active faults in crystalline rock. *Geology*, **23**(8), 683-686. https://doi.org/10.1130/0091-7613(1995)023<0683:FFAPAF>2.3.CO;2

Bense, V. F., T. Gleeson, S. E. Loveless, O. Bour, & J. Scibek (2013). Fault zone hydrogeology. *Earth-Science Reviews*, **127**, 171-192. https://doi.org/10.1016/j.earscirev.2013.09.008

Ben-Zion, Y., & Z. Shi. (2005). Dynamic rupture on a material interface with spontaneous generation of plastic strain in the bulk. *Earth and Planetary Science Letters* 236(1-2), 486-496. https://doi.org/10.1016/j.epsl.2005.03.025

Brace, W. F., B. W. Paulding Jr, & C. H. Scholz (1966). Dilatancy in the fracture of crystalline rocks. *Journal of Geophysical Research*, **71**(16), 3939-3953.https://doi.org/10.1029/JZ071i016p03939

Brace, W. F. (1978). A note on permeability changes in geologic material due to stress. *Pure and applied geophysics*, **116**, 627-633. https://doi.org/10.1007/BF00876529

Brantut, N. (2020). Dilatancy-induced fluid pressure drop during dynamic rupture: Direct experimental evidence and consequences for earthquake dynamics. *Earth and Planetary Science Letters*, **538**, 116179. https://doi.org/10.1016/j.epsl.2020.116179

Boore, D. M., & W. B. Joyner (1997). Site amplifications for generic rock sites. *Bulletin of the seismological society of America*, **87**(2), 327-341. https://doi.org/10.1785/BSSA0870020327

Brocher, T. M. (2005). Empirical relations between elastic wavespeeds and density in the Earth's crust. *Bulletin of the seismological Society of America,* **95**(6), 2081-2092. https://doi.org/10.1785/0120050077

Caine, J. S., & C. B. Forster (1999). Fault zone architecture and fluid flow: Insights from field data and numerical modeling. *Geophysical Monograph-American Geophysical Union*, **113**, 101-128.

Cappa, F. (2009). Modelling fluid transfer and slip in a fault zone when integrating heterogeneous hydromechanical characteristics in its internal structure. *Geophysical Journal International*, **178**(3), 1357-1362. https://doi.org/10.1111/j.1365-246X.2009.04291.x

Chernyshev, S. N., & W. R. Dearman (1991). *Rock Fractures*, Butterworth-Heinemann, London.

Chen, X., Y. Li, L. Tong, D. Geng, Z. Dong & P. Yang (2023). Elastoplastic Damage Behavior of Rocks: A Case Study of Sandstone and Salt Rock. *Rock Mech Rock Eng* **56**, 5621–5634. https://doi.org/10.1007/s00603-023-03349-5

Cocco, M., & J. R. Rice (2002). Pore pressure and poroelasticity effects in Coulomb stress analysis of earthquake interactions. *Journal of Geophysical Research: Solid Earth*, **107**(B2), ESE-2. https://doi.org/10.1029/2000JB000138

Cochran, E. S., Y.-G. Li, P. M. Shearer, S. Barbot, Y. Fialko & J. E. Vidale (2009). Seismic and geodetic evidence for extensive, long-lived fault damage zones. *Geology*, **37**(4), 315–318. https://doi.org/10.1130/G25306A.1

Cox, S. F. (2010). The application of failure mode diagrams for exploring the roles of fluid pressure and stress states in controlling styles of fracture-controlled permeability enhancement in faults and shear zones. *Geofluids*, **10**(1–2), 217–233. https://doi.org/10.1111/j.1468-8123.2010.00281.x

Cox, S. F. (1999). Deformational controls on the dynamics of fluid flow in mesothermal gold systems. *Geological Society, London, Special Publications*, **155**(1), 123-140. https://doi.org/10.1144/GSL.SP.1999.155.01.10

Dor, O., T. K. Rockwell, & Y. Ben-Zion (2006). Geological observations of damage asymmetry in the structure of the San Jacinto, San Andreas and Punchbowl faults in Southern California: A possible indicator for preferred rupture propagation direction. *Pure and Applied Geophysics*, **163**, 301-349. https://doi.org/10.1007/s00024-005-0023-9

Drucker, D. C., & W. Prager (1952), Soil mechanics and plastic analysis or limit design, *Quarterly of applied mathematics*, **10**, 157–165.

Dunham, E. M., D. Belanger, L. Cong, J. E. Kozdon (2011). Earthquake Ruptures with Strongly Rate-Weakening Friction and Off-Fault Plasticity, Part 1: Planar Faults. *Bulletin of the Seismological Society of America*, 101 (5): 2296–2307. https://doi.org/10.1785/0120100075

Evans, J. P., C. B. Forster, & J. V. Goddard (1997). Permeability of fault-related rocks, and implications for hydraulic structure of fault zones. *Journal of structural Geology*, **19**(11), 1393-1404. https://doi.org/10.1016/S0191-8141(97)00057-6

Farrell, N. J. C., D. Healy & C. W. Taylor (2014). Anisotropy of permeability in faulted porous sandstones. *Journal of structural geology*, **63**, 50-67. https://doi.org/10.1016/j.jsg.2014.02.008

Faulkner, D. R., & P. J. Armitage (2013). The effect of tectonic environment on permeability development around faults and in the brittle crust. *Earth and Planetary Science Letters*, **375**, 71-77. https://doi.org/10.1016/j.epsl.2013.05.006

Fleeger, G. M. (1999). Hydrologic effects of the Pymatuning earthquake of September 25, 1998, in northwestern Pennsylvania, *US Department of the Interior, US Geological Survey*, **99**(4170).

Freeze, R., & J. A. Cherry (1979). *Groundwater*. Prentice-Hall.

Frery, E., J.-P. Gratier, N. Ellouz-Zimmerman, C. Loiselet, J. Braun, P. Deschamps, D. Blamart, B. Hamelin, & R. Swennen (2015). Evolution of fault permeability during episodic fluid circulation: Evidence for the effects of fluid–rock interactions from travertine studies (Utah–USA), *Tectonophysics*, **651–652**, 121-137. https://doi.org/10.1016/j.tecto.2015.03.018.

Gabriel, A.-A., J.-P. Ampuero, L. A. Dalguer, and P. M. Mai (2013), Source properties of dynamic rupture pulses with off-fault plasticity, Journal of Geophysical Research, 118, 4117–4126, doi:10.1002/jgrb.50213

Ge, S., & S. C. Stover (2000). Hydrodynamic response to strike- and dip-slip faulting in a half-space. *Journal of Geophysical Research*, **105**(B11), 25513–25524.

https://doi.org/10.1029/2000JB900233

Grawinkel, A., & B. Stöckhert (1997). Hydrostatic pore fluid pressure to 9 km depth-Fluid inclusion evidence from the KTB deep drill hole. *Geophysical Research Letters*, **24**(24), 3273-3276. https://doi.org/10.1029/97GL03309

Gudmundsson, A. (2000). Active fault zones and groundwater flow. *Geophysical Research Letters*, **27**(18), 2993–2996. https://doi.org/10.1029/1999GL011266

Guéguen, Y., & A. Schubnel (2003). Elastic wave velocities and permeability of cracked rocks *Tectonophysics*, 370(1-4), 163-176. https://doi.org/10.1016/S0040-1951(03)00184-7

Grueschow, E., O. Kwon, I. G. Main, & J. W. Rudnicki (2003). Observation and modeling of the suction pump effect during rapid dilatant slip. *Geophysical Research Letters*, **30**(5). https://doi.org/10.1029/2002GL015905

Heap, M. J., D. R. Faulkner, P. G. Meredith, & S. Vinciguerra (2010). Elastic moduli evolution and accompanying stress changes with increasing crack damage: implications for stress changes around fault zones and volcanoes during deformation. *Geophysical Journal International*, *183*(1), 225-236. https://doi.org/10.1111/j.1365-246X.2010.04726.x

Hirakawa, E., & S. Ma (2016), Dynamic fault weakening and strengthening by gouge compaction and dilatancy in a fluid-saturated fault zone, J. Geophys. Res. Solid Earth, 121, 5988–6008, https://doi.org/10.1002/2015JB012509.

Hosono, T., C. Yamada, T. Shibata, Y. Tawara, C.-Y. Wang, M. Manga, M., et al. (2019). Coseismic groundwater drawdown along crustal ruptures during the 2016 Mw 7.0 Kumamoto earthquake. *Water Resources Research*, **55**(7), 5891–5903. https://doi.org/10.1029/2019WR024871

Hosono, T., C. Yamada, M. Manga, C.-Y. Wang, & M. Tanimizu (2020). Stable isotopes show that earthquakes enhance permeability and release water from mountains. *Nature Communications*, **11**(1), 2776. https://doi.org/10.1038/s41467-020-16604-y

Hosono, T. & Y. Masaki (2020). Post-seismic hydrochemical changes in regional groundwater flow systems in response to the 2016 Mw 7.0 Kumamoto earthquake. Journal of Hydrology, 580, 124340. https://doi.org/10.1016/j.jhydrol.2019.124340.

Hung, R.-J., M. Weingarten, & M. Manga (2024). Very-near-field coseismic fault pressure drop and delayed postseismic cross-fault flow induced by fault damage from the 2018 M6. 3 Hualien, Taiwan earthquake. *Journal of Geophysical Research: Solid Earth*, **129**(10), e2024JB029188. https://doi.org/10.1029/2024JB029188

Hung, R.-J. (2025). Earthquake_dynamic_rupture_code-v3, Mendeley Data, V2 [software]. https://data.mendeley.com/datasets/gf5pbn39gf/2

Ingebritsen, S. E., & M. Manga (2019). Earthquake hydrogeology. *Water Resources Research*, 55, 5212–5216. https://doi.org/10.1029/2019WR025341

Johri, M., E. M. Dunham, M. D. Zoback & Z. Fang(2014). Predicting fault damage zones by modeling dynamic rupture propagation and comparison with field observations. *Journal of Geophysical Research*, **119**(2), 1251-1272. https://doi.org/10.1002/2013JB010335

Jónsson, S., P. Segall, R. Pedersen, & G. Björnsso (2003). Post-earthquake ground movements correlated to pore-pressure transients. *Nature*, **424**(6945), 179–183. https://doi.org/10.1038/nature01776

Koizumi, N., Y. Kano, Y. Kitagawa, T. Sato, M. Takahashi, S. Nishimura, & R. Nishida (1996). Groundwater Anomalies Associated with the 1995 Hyogo-ken Nanbu Earthquake. *Journal of Physics of the Earth*, 44, 373-380. https://doi.org/10.4294/jpe1952.44.373

Lai, W. C., N. Koizumi, N. Matsumoto, Y. Kitagawa, C. W. Lin, C. L. Shieh, & Y.-P. Lee (2004). Effects of seismic ground motion and geological setting on the coseismic groundwater level changes caused by the 1999 Chi-Chi earthquake, Taiwan. *Earth, planets and space*, *56*, 873-880.

Langevin, C. D., Hughes, J. D., Banta, E. R., Niswonger, R. G., Panday, S., & Provost, A. M. (2017). Documentation for the MODFLOW 6 groundwater flow model [software]. U.S. Geological Survey Techniques and Methods, 197. https://doi.org/10.3133/tm6A55

Leonard, M. (2014). Self-consistent earthquake fault-scaling relations: Update and extension to stable continental strike-slip faults. *Bulletin of the Seismological Society of America*, **104**(6), 2953-2965. https://doi.org/10.1785/0120140087

Liao, X., C.-Y Wang, & C.-P. Liu (2015). Disruption of groundwater systems by earthquakes. *Geophysical Research Letters*, *42*(22), 9758-9763. https://doi.org/10.1002/2015GL066394

Liu, Z., & B. Duan (2014). Dynamics of Parallel Strike-Slip Faults with Pore Fluid Pressure Change and Off-Fault Damage. *Bulletin of the Seismological Society of America*, 104 (2), 780–792. https://doi.org/10.1785/0120130112

Liu, R., B. Li, L. Yu, Y. Jiang, & H. Jing (2018). A discrete-fracture-network fault model revealing permeability and aperture evolutions of a fault after earthquakes. *International Journal of Rock Mechanics and Mining Sciences*, *107*, 19-24.

https://doi.org/10.1016/j.ijrmms.2018.04.036

Lockner, D. A., & J. D. Byerlee (1994). Dilatancy in hydraulically isolated faults and the suppression of instability. *Geophysical Research Letters*, *21*(22), 2353-2356.

https://doi.org/10.1029/94GL02366

Lockner, D. A., H. Tanaka, H. Ito, R. Ikeda, K. Omura, & H. Naka (2009). Geometry of the Nojima fault at Nojima-Hirabayashi, Japan—I. A simple damage structure inferred from borehole core permeability. *Pure and Applied Geophysics*, **166**(10–11), 1649–1667.

https://doi.org/10.1007/s00024-009-0515-0

Ma, S. (2008). A physical model for widespread near-surface and fault zone damage induced by earthquakes. *Geochemistry, Geophysics, Geosystems*, *9*(11).

https://doi.org/10.1029/2008GC002231

Ma, S. (2012). A self-consistent mechanism for slow dynamic deformation and tsunami generation for earthquakes in the shallow subduction zone. *Geophysical Research Letters*, **39**(11).

https://doi.org/10.1029/2012GL051854

Ma, S. & D. J. Andrews (2010). Inelastic off-fault response and three-dimensional dynamics of earthquake rupture on a strike-slip fault. *Journal of Geophysical Research*, **115**(B4). https://doi.org/10.1029/2009JB006382.

Ma, S. (2022). Dynamic off-fault failure and tsunamigenesis at strike-slip restraining bends: Fully-coupled models of dynamic rupture, ocean acoustic waves, and tsunami in a shallow bay. *Tectonophysics*, *838*, 229496. https://doi.org/10.1016/j.tecto.2022.229496

Manga, M., I. Beresnev, E. E. Brodsky, J. E. Elkhoury, D. Elsworth, S. E. Ingebritsen, et al. (2012). Changes in permeability caused by transient stresses: Field observations, experiments, and mechanisms. *Reviews of Geophysics*, **50**(2). https://doi.org/10.1029/2011RG000382

Manning, C. E., & S. E. Ingebritsen (1999). Permeability of the continental crust: Implications of geothermal data and metamorphic systems. *Reviews of Geophysics*, *37*, 127-150. https://doi.org/10.1029/1998RG900002

Matthäi, S. K. & G. Fischer (1996). Quantitative modeling of fault-fluid-discharge and fault-dilation-induced fluid-pressure variations in the seismogenic zone. *Geology*, **24**(2), 183–186. https://doi.org/10.1130/0091-7613(1996)024<0183:QMOFFD>2.3.CO;2

Miao, M., S. Zhu, Y. Chang, J. Yuan, R. Wang, & P. Han (2021). Spatiotemporal evolution of pore pressure changes and Coulomb failure stress in a poroelastic medium for different faulting regimes. *Earth and Space Science*, *8*(11), e2021EA001837. https://doi.org/10.1029/2021EA001837

Morton, N., G. H. Girty, & T. K. Rockwell (2012). Fault zone architecture of the San Jacinto fault zone in Horse Canyon, southern California: A model for focused post-seismic fluid flow and heat transfer in the shallow crust. *Earth and Planetary Science Letters*, *329*, 71-83. https://doi.org/10.1016/j.epsl.2012.02.013

Muir-Wood, R., & G. C. King (1993). Hydrological signatures of earthquake strain. *Journal of Geophysical Research: Solid Earth*, *98*(B12), 22035-22068. https://.doi.org/10.1029/93JB02219.

Nakagawa, K., Z.-Q. Yu, R. Berndtsson, & T. Hosono (2020). Temporal characteristics of groundwater chemistry affected by the 2016 Kumamoto earthquake using self-organizing maps, Journal of Hydrology, 582, 124519. https://doi.org/10.1016/j.jhydrol.2019.124519.

Nur, A., & J. R. Booker (1972). Aftershocks caused by pore fluid flow? *Science*, *175*(4024), 885-887. https://doi.org/10.1126/science.175.4024.885

Peach, C. J., & C. J. Spiers (1996). Influence of crystal plastic deformation on dilatancy and permeability development in synthetic salt rock. *Tectonophysics*, *256*(1-4), 101-128. https://doi.org/10.1016/0040-1951(95)00170-0

Proctor, B., D. A. Lockner, B. D. Kilgore, T. M. Mitchell & N. M. Beeler (2020). Direct evidence for fluid pressure, dilatancy, and compaction affecting slip in isolated faults. *Geophysical Research Letters*, **47**(16), e2019GL086767. https://doi.org/10.1029/2019GL086767

Quilty, E. G., & E. A. Roeloffs (1997). Water-level changes in response to the 20 December 1994 earthquake near Parkfield, California. *Bulletin of the Seismological Society of America*, 87(2), 310–317. https://doi.org/10.1785/BSSA0870020310

Rempe, M., T. M. Mitchell, J. Renner, S. A. F. Smith, A. Bistacchi, & G. Di Toro (2018). The relationship between microfracture damage and the physical properties of fault-related rocks: The Gole Larghe Fault Zone, Italian Southern Alps. Journal of Geophysical Research: Solid Earth, 123, 7661–7687. https://doi. org/10.1029/2018JB015900

Rice, J. R. (1992). Chapter 20 fault stress states, pore pressure distributions, and the weakness of the San Andreas Fault. In B. Evans, & T. Wong (Eds.), *International Geophysics* (Vol. 51, pp. 475–503). Academic Press. https://doi.org/10.1016/S0074-6142(08)62835-1

Rice, J. R. (2006). Heating and weakening of faults during earthquake slip. *Journal of Geophysical Research: Solid Earth*, *111*(B5). https://doi.org/10.1029/2005JB004006

Roeloffs, E. (1996). Poroelastic techniques in the study of earthquake-induced hydrologic phenomena. *Advances in Geophysics*, **37**, 133–195. https://doi.org/10.1016/S0065-2687(08)60270-8

Rojstaczer, S., S. Wolf, & R. Michel (1995). Permeability enhancement in the shallow crust as a cause of earthquake-induced hydrological changes. *Nature*, **373**(6511), 237–239. https://doi.org/10.1038/373237a0

Ross, Z. E., E. S. Cochran, D. T. Trugman, & J. D. Smith (2020). 3D fault architecture controls the dynamism of earthquake swarms. *Science*, *368*(6497), 1357-1361. https://doi.org/10.1126/science.abb0779

Roten, D., K. B. Olsen, & S. M. Day (2017). Off-fault deformations and shallow slip deficit from dynamic rupture simulations with fault zone plasticity. *Geophysical Research Letters*, **44**(15), 7733-7742.

Rutter, H. K., S. C. Cox, N. F. D. Ward, & J. J. Weir (2016). Aquifer permeability change caused by a near-field earthquake, Canterbury, New Zealand. *Water Resources Research*, **52**(11), 8861–8878. https://doi.org/10.1002/2015WR018524

Scholz, C. H., L. R. Sykes, & Y. P. Aggarwal (1973). Earthquake Prediction: A Physical Basis: Rock dilatancy and water diffusion may explain a large class of phenomena precursory to earthquakes. *Science*, **181**(4102), 803-810. https://doi.org/10.1126/science.181.4102.803

Scholz, C. H. (2000). Evidence for a strong San Andreas fault. *Geology*, **28**(2), 163-166. https://doi.org/10.1130/0091-7613(2000)28<163:EFASSA>2.0.CO;2

Scholz, C. H. (2019). *The mechanics of earthquakes and faulting*. University of Cambridge. https://doi.org/10.1017/9781316681473

Shi, Z., G. Wang & C. Liu (2013). Co-seismic groundwater level changes induced by the May 12, 2008 Wenchuan earthquake in the near field. *Pure and Applied Geophysics*, **170**, 1773-1783. https://doi.org/10.1007/s00024-012-0606-1

Sibson, R. H. (1983). Continental fault structure and the shallow earthquake source. *Journal of the Geological Society*, **140**(5), 741-767. https://doi.org/10.1144/gsjgs.140.5.074

Sibson, R. H. (1989). Earthquake faulting as a structural process. Journal of Structural Geology, **11**, 1-14. https://doi.org/10.1016/0191-8141(89)90032-1

Sibson, R. H. (1990). Conditions for fault-valve behaviour. *Geological Society, London, Special Publications*, **54**(1), 15-28. https://doi.org/10.1144/gsjgs.140.5.0741

Sibson, R. H. (1994). Crustal stress, faulting and fluid flow. *Geological Society, London, Special Publications*, **78**(1), 69-84. https://doi.org/10.1144/GSL.SP.1994.078.01.07

Sibson, R. H. (1996). Structural permeability of fluid-driven fault-fracture meshes. *Journal of Structural geology*, **18**(8), 1031-1042. https://doi.org/10.1016/0191-8141(96)00032-6

Shmonov, V. M., V. M. Vitiovtova., A. V. Zharikov, & A. A. Grafchikov (2003). Permeability of the continental crust: implications of experimental data. *Journal of Geochemical Exploration*, *78*, 697-699. https://doi.org/10.1016/S0375-6742(03)00129-8

Tadokoro, K., M. Ando & K. Y. Nishigami (2000). Induced earthquakes accompanying the water injection experiment at the Nojima fault zone, Japan: seismicity and its migration. *Journal of Geophysical Research: Solid Earth*, **105**(B3), 6089-6104. https://doi.org/10.1029/1999JB900416

Taufiqurrahman, T., A.-A. Gabriel, D. Li, T. Ulrich, B. Li et al. (2023). Dynamics, interactions and delays of the 2019 Ridgecrest rupture sequence. *Nature,* **618**, 308–315. https://doi.org/10.1038/s41586-023-05985-x

Tokunaga, T. (1999). Modeling of earthquake-induced hydrological changes and possible permeability enhancement due to the 17 January 1995 Kobe earthquake, Japan. *Journal of Hydrology,* **223**, 221–229. https://doi.org/10.1016/S0022-1694(99)00124-9

Townend, J. & M. Zoback (2000). How faulting keeps the crust strong. *Geology*, 28(5), 399–402. https://doi.org/10.1130/0091-7613(2000)28<399:HFKTCS>2.0.CO;2

Viesca, R. C., E. L. Templeton & J. R. Rice (2008). Off-fault plasticity and earthquake rupture dynamics: 2. Case of saturated off-fault materials. *Journal of Geophysical Research*, **113**(B9), B09307. https://doi.org/10.1029/2007JB005530

Wang, H. F. (2000). *Theory of linear poroelasticity with applications to geomechanics and hydrogeology*. Princeton University Press.

Wang, C.-Y., M. Manga, D. Dreger & A. Wong (2004a), Streamflow increase due to rupturing of hydrothermal reservoirs: Evidence from the 2003 San Simeon, California, Earthquake, Geophysical Research Letters, **31**, L10502, https://doi.org/10.1029/2004GL020124.

Wang, C.-Y., C. H. Wang & M. Manga (2004b). Coseismic release of water from mountains: Evidence from the 1999 (Mw= 7.5) Chi-Chi, Taiwan, earthquake. *Geology*, **32**(9), 769-772. https://doi.org/10.1130/G20753.1

Wang, C.-Y., & M. Manga (2010). Hydrologic responses to earthquakes and a general metric. Geofluids, **10**(1-2), 206-216. https://doi.org/10.1111/j.1468-8123.2009.00270.x

Wang, C.-Y., & M. Manga (2015). New streams and springs after the 2014 Mw6. 0 South Napa earthquake. *Nature Communications*, **6**(1), 7597. https://doi.org/10.1038/ncomms8597

Wang, C.-Y., & M. Manga. (2021). Water and earthquakes, *Spring Nature*. https://doi.org/10.1007/978-3-030-64308-9

Wibberley, C. A. J., & T. Shimamoto (2003). Internal structure and permeability of major strike-slip fault zones: the Median Tectonic Line in Mie Prefecture, Southwest Japan. *Journal of Structural Geology*, **25**(1), 59-78. https://doi.org/10.1016/S0191-8141(02)00014-7.

Woodcock, N. H., & M. Fischer (1986). Strike-slip duplexes. *Journal of structural geology*, **8**(7), 725-735. https://doi.org/10.1016/0191-8141(86)90021-0

Xu, S., Y. Ben-Zion & J. P. Ampuero (2012). Properties of inelastic yielding zones generated by in-plane dynamic ruptures—I. Model description and basic results. *Geophysical Journal International*, **191**(3), 1325-1342. https://doi.org/10.1111/j.1365-246X.2012.05679.x

Wu, G., L. Gao, Y. Zhang, C. Ning & E. Xie (2019). Fracture attributes in reservoir-scale carbonate fault damage zones and implications for damage zone width and growth in the deep subsurface. *Journal of Structural Geology*, *118*, 181-193.
https://doi.org/10.1016/j.jsg.2018.10.008

Yao, L., S. Ma & G. Di Toro (2023). Coseismic fault sealing and fluid pressurization during earthquakes. *Nature Communications*, **14**(1), 1136. https://doi.org/10.1038/s41467-023-36839-9

Zhang, X., & D. J. Sanderson (1996). Numerical modelling of the effects of fault slip on fluid flow around extensional faults. *Journal of Structural Geology*, **18**(1), 109-119. https://doi.org/10.1016/0191-8141(95)00086-S

Zhang, Y., P. Schaubs, C. Zhao, A. Ord, B. Hobbs, A. Barnicoat (2008). Fault-related dilation, permeability enhancement, fluid flow and mineral precipitation patterns: numerical models The Internal Structure of Fault Zones: Implications for Mechanical and Fluid-flow Properties, *The Geological Society of London*. https://doi.org/10.1144/SP299.15

Zhao, X. G., & M. Cai (2010). A mobilized dilation angle model for rocks. *International Journal of Rock Mechanics and Mining Sciences*, **47**(3), 368-384. https://doi.org/10.1016/j.ijrmms.2009.12.007

Zoback, M. D., & J. Townend (2001). Implications of hydrostatic pore pressures and high crustal strength for the deformation of intraplate lithosphere. *Tectonophysics*, **336**(1-4), 19-30. https://doi.org/10.1016/S0040-1951(01)00091-9

Zhou, X., & T. J. Burbey; (2014). Pore-Pressure Response to Sudden Fault Slip for Three Typical Faulting Regimes. *Bulletin of the Seismological Society of America*, **104**(2), 793–808. https://doi.org/10.1785/0120130139

Zhu, W., & T. F. Wong (1999). Network modeling of the evolution of permeability and dilatancy in compact rock. *Journal of Geophysical Research*, **104**(B2), 2963-2971. https://doi.org/10.1029/1998JB900062

Zhu, W., K. L. Allison, E. M. Dunham, & Y. Yang (2020). Fault valving and pore pressure evolution in simulations of earthquake sequences and aseismic slip. *Nature communications*, **11**(1), 4833.

Zoback, M. D., & J. Townend (2001). Implications of hydrostatic pore pressures and high crustal strength for the deformation of intraplate lithosphere. *Tectonophysics*, **336**(1-4), 19-30. https://doi.org/10.1016/S0040-1951(01)00091-9

**Table**

| Earthquake (Ref) | Observed Distance from Fault | Observed Water Level / Pore Pressure Change | Recovery Behavior | Notes on Relevance to This Study | Reference |
|---|---|---|---|---|---|
| **Hualien, Taiwan 2018** | 0.18 km – 2 km | 2 – 15 m drawdown | Weeks–months | Our modeled generic case produces 10–40 m drawdown at similar distances; few meters at ~2 km | Hung et al. (2024) |
| **Kumamoto, Japan 2016** | <1 km | ~5 m drawdown; some recharge above baseline | Recovery to above baseline within weeks | Consistent with modeled patterns of localized drawdown and over-recharge | Hosono et al., (2019); (2020) |
| **Kobe, Japan 1995** | Possibly <1 km | Up to 70 m decline (over ~3 months) | Prolonged, incomplete recovery | Extreme case showing drawdowns comparable to our largest modeled near-fault values | Tokunaga, (1999) |

| Chi-Chi, Taiwan 1999 | km-scale near rupture | 5–10 m drawdowns near fault rupture and several meters of increase in the footwall | Recovery in months | Consistent with moderate modeled drawdowns at 1–2 km. Groundwater recharge from the mountain range is not modeled | Wang et al., (2004b); Lai et al., (2004) |
|---|---|---|---|---|---|
| Wenchuan, China 2008 | Near-field 21 km from fault) | 13 m drawdowns | Months–years | The drawdown is more likely triggered by local damage from seismic wave, not by fault-zone damage. | Liao et al. (2015) |
| Canterbury, New Zealand 2010 | Near-fault wells | Water level increase (permeability clogging) | Weeks | Contrasts with our model (which lacks permeability healing) | Rutter et al., (2016) |

Table 1 Observed near-fault coseismic groundwater response versus model performance.

**Figures**

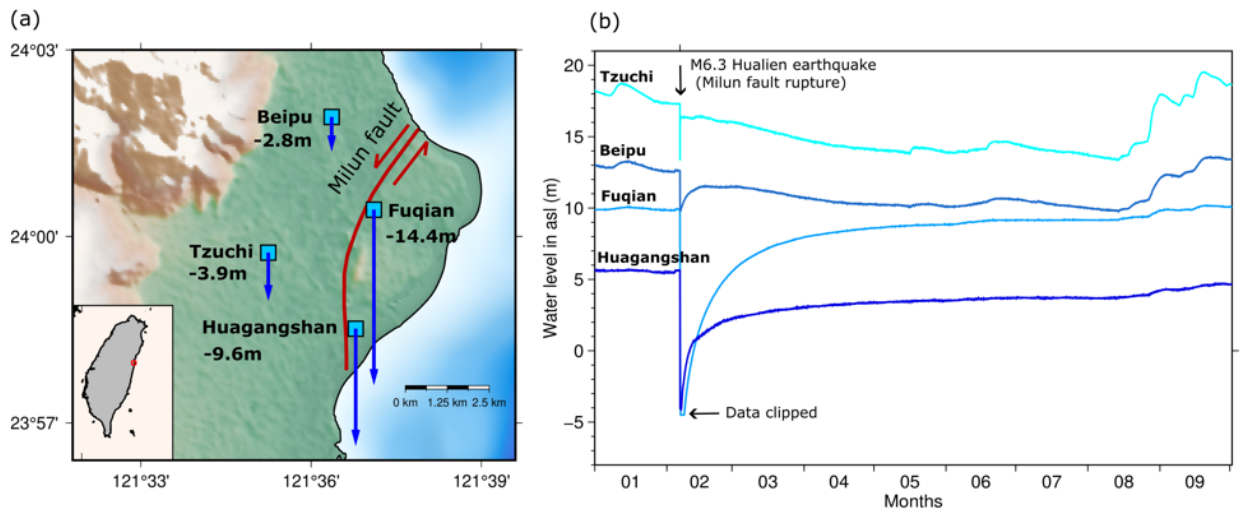

Figure 1. Near-fault hydrologic observation during the 2018 Hualien, Taiwan earthquake collected from the continuous groundwater wells (depth 77-190m). (a) map showing station locations and the ruptured strike-slip Milun fault which is responsible for the earthquake. Blue arrows and numbers (reduction negative) indicate the maximum magnitudes of water-level change. (b) Time series data for the groundwater levels in response to the earthquake (after Hung et al., 2024).

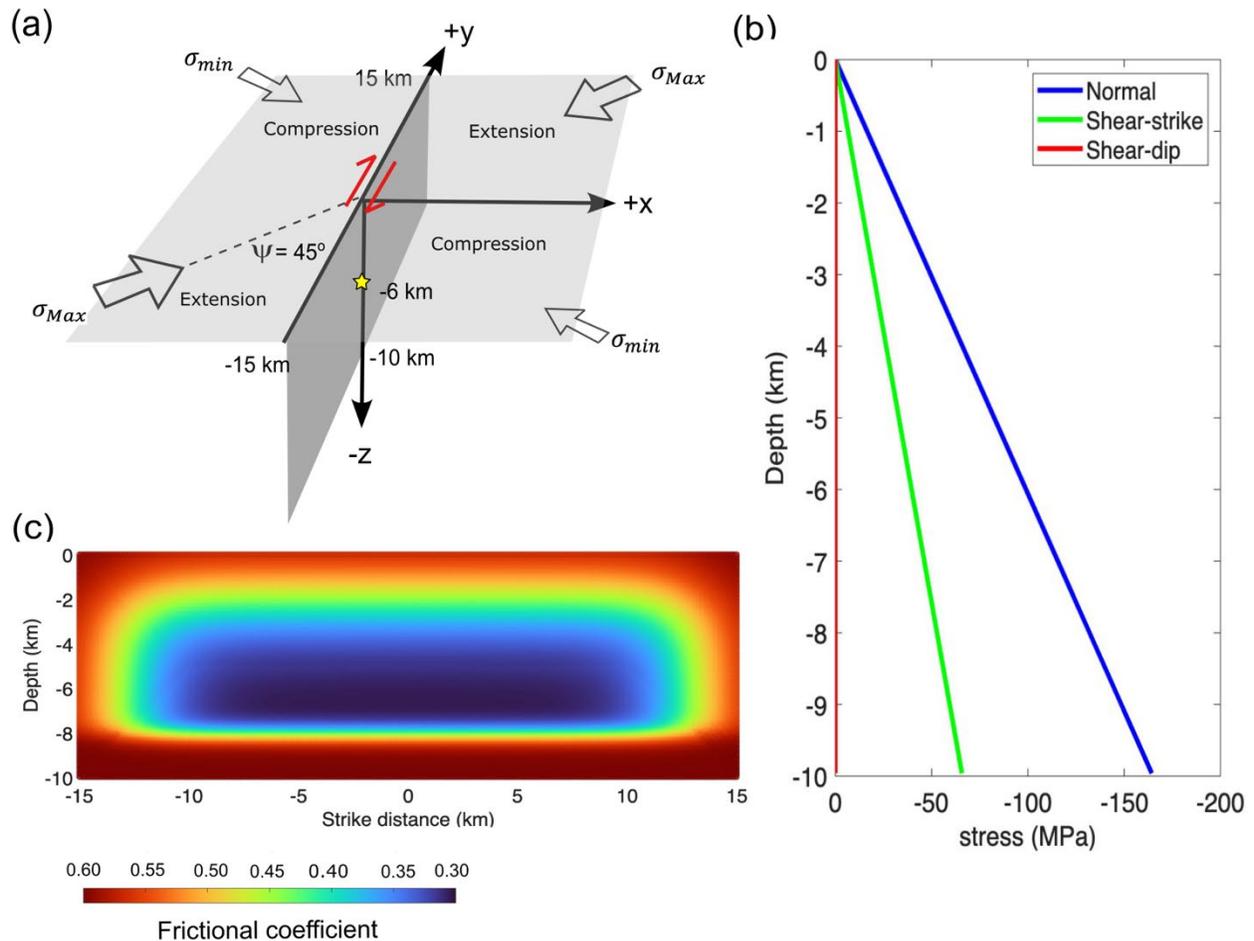

Figure 2. Initial setting of the fault model for the earthquake dynamic rupture simulation. (a) fault dimension and geometry (b) initial shear and normal stresses in the model (c) initial dynamic frictional coefficients on the fault plane. The initial static friction = 0.6 is a constant throughout the fault plane.

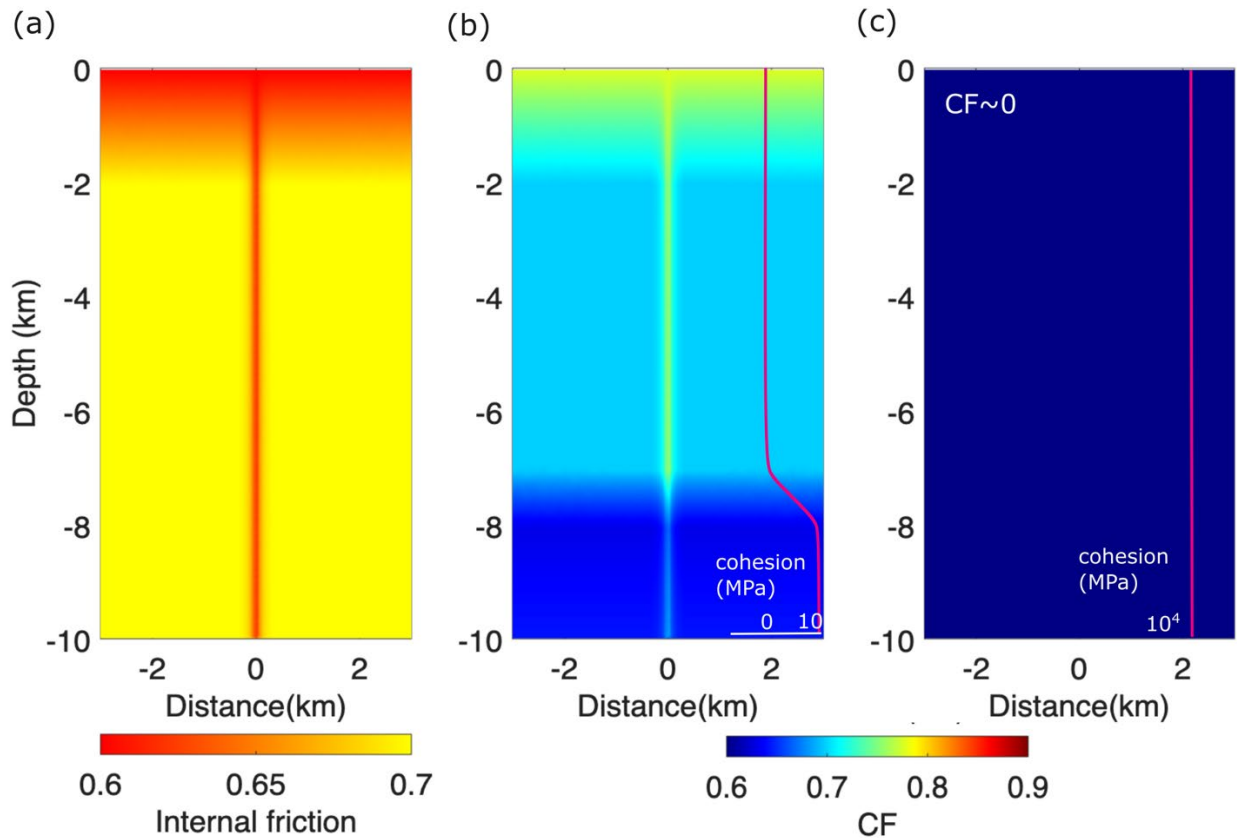

Figure 3. Material properties are prescribed as the initial condition for the off-fault damage (a) internal frictions, (b) closeness-of-failure (CF) for the inelastic model, and (c) CF for the elastic model (CF~0). Red lines in (b) & (c) indicate the cohesion value.

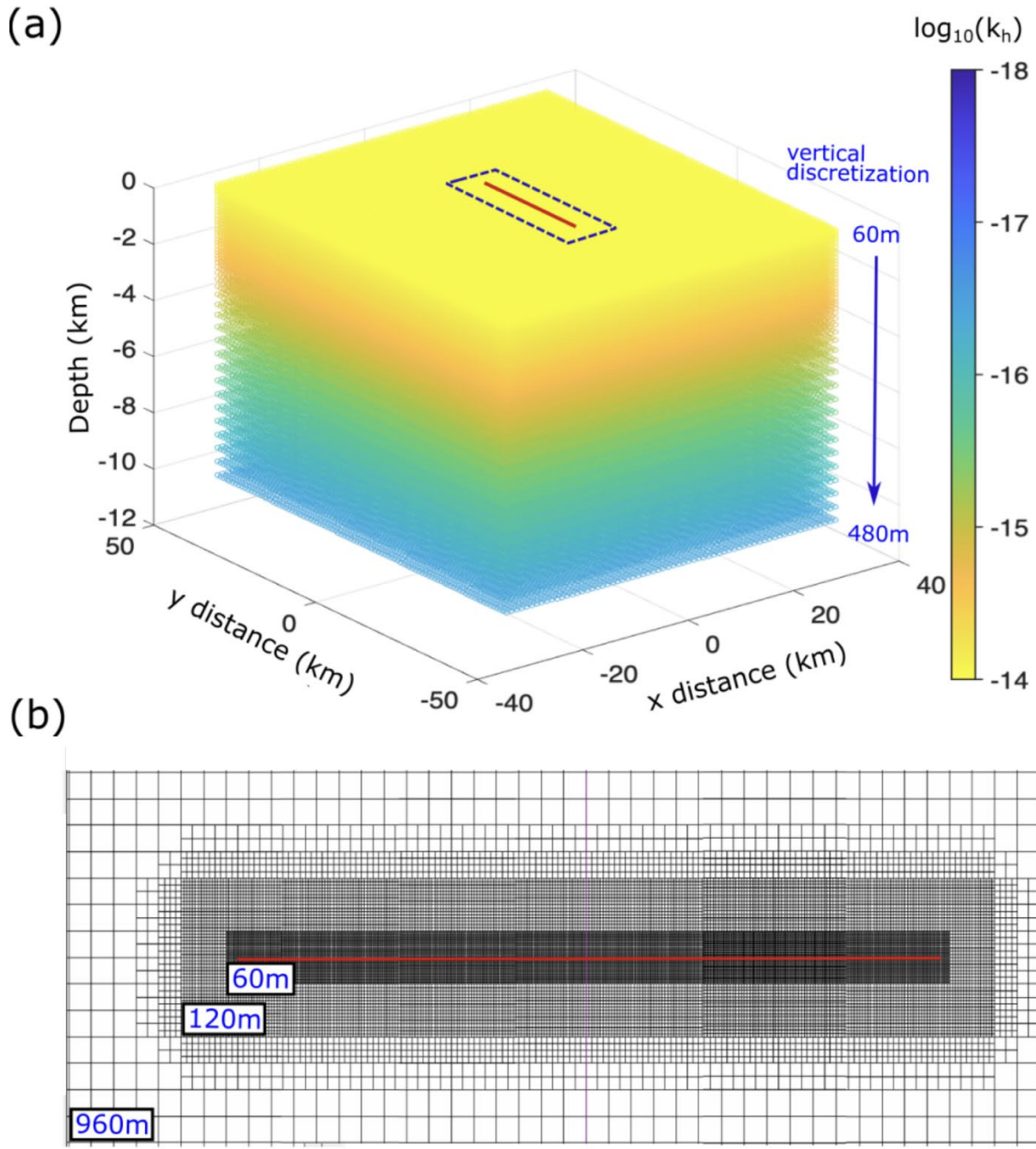

Figure 4. Model structure for the postseismic hydrologic simulation. (a) 3-D model dimension and the initial permeability setting. Dashed box highlights the fault-zone grids plotted in (b). Red line on the model top surface indicates the modeled fault zone. (b) Zoom-in plot of the unstructured grids around the fault. Blue numbers indicate the different grid sizes. Red line indicates the fault trace.

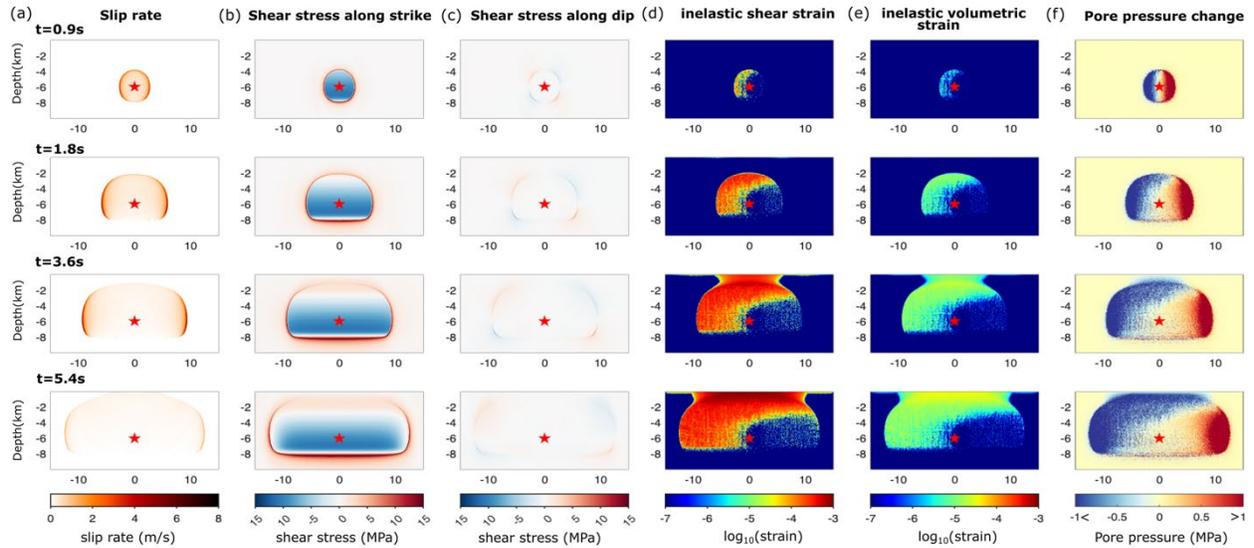

Figure 5. Snapshots of dynamic rupture parameters on the fault plane for (a) slip rate (b) along-strike shear stress (c) along-dip shear stress (d) inelastic shear strain (e) inelastic volumetric strain and (f) pore pressure change. See also Movie S1.

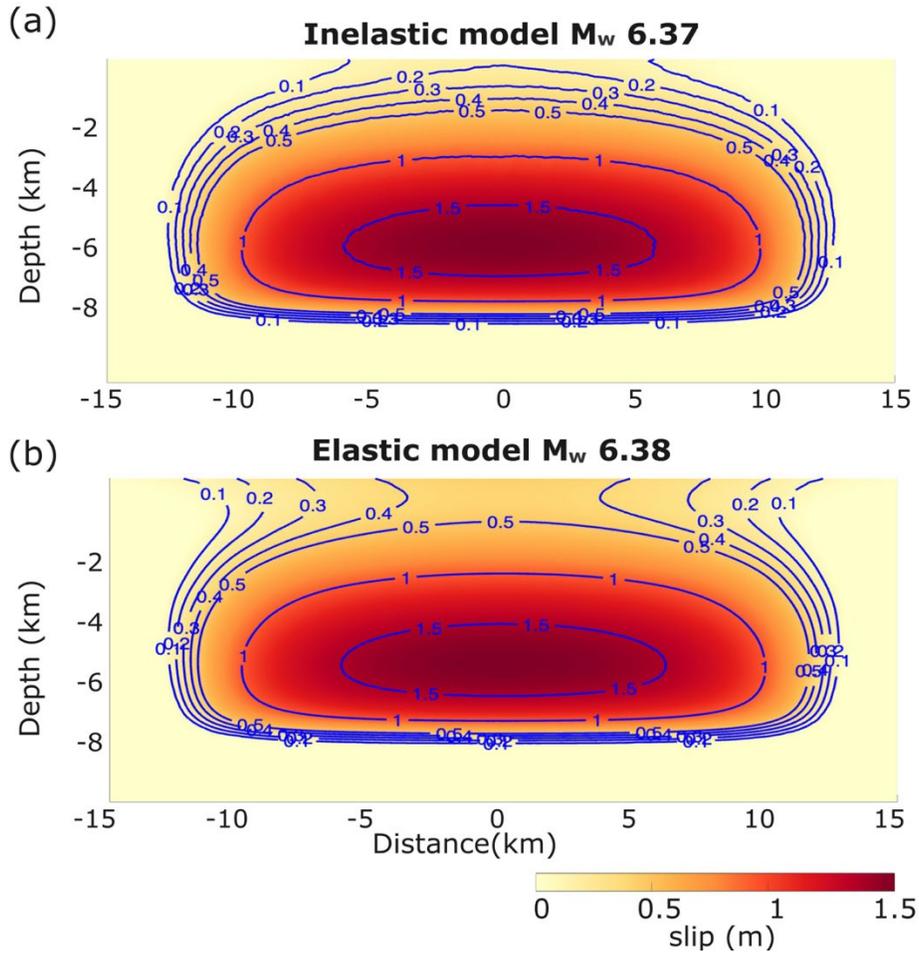

Figure 6. Slip distribution generated by the earthquake dynamic rupture modeling between (a) inelastic model and (b) elastic model. Contours indicate the slip values. Moment magnitude is calculated via $(\log_{10}M_0 - 9.1)/1.5$, where $M_0$ is seismic moment.

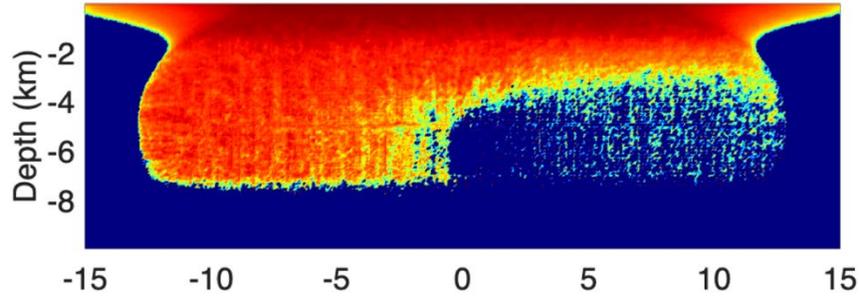

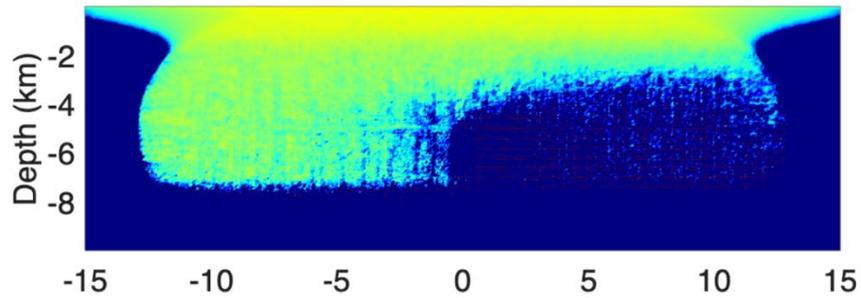

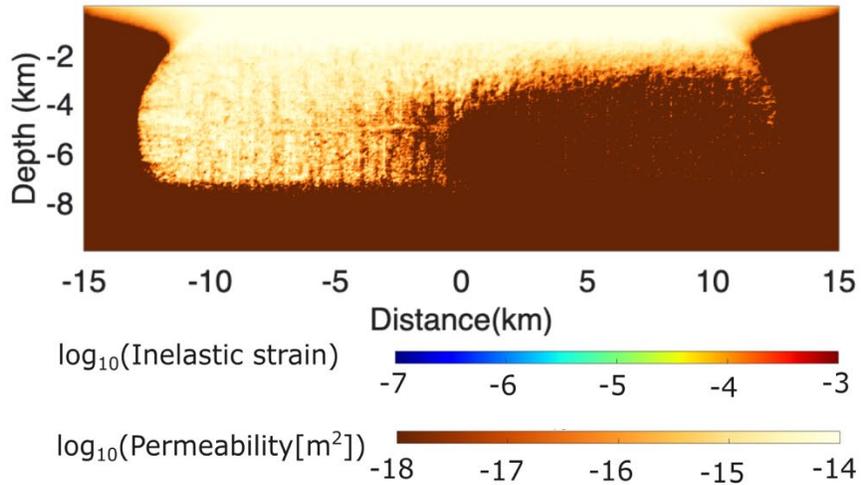

Figure 7. Coseismic inelastic strains and the corresponding permeability distributions on the nearest slice (x= -30 m) from the modeled fault plane (x = 0 m) obtained from the earthquake dynamic rupture model. Negative x is in dilation quadrant (see Fig 2a). (a) inelastic shear strain, (b) inelastic volumetric strain, and (c) converted permeability.

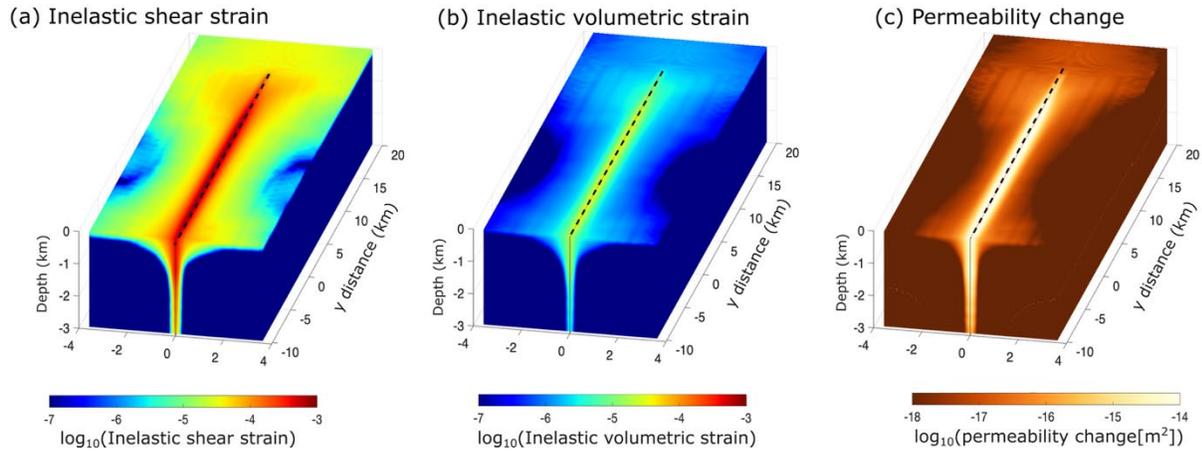

Figure 8. Inelastic strain and permeability distributions in the off-fault regions plotted in 3-D, showing damage variations both inside and outside the fault. (a) inelastic shear strain (b) inelastic volumetric strain, and (c) converted permeability.

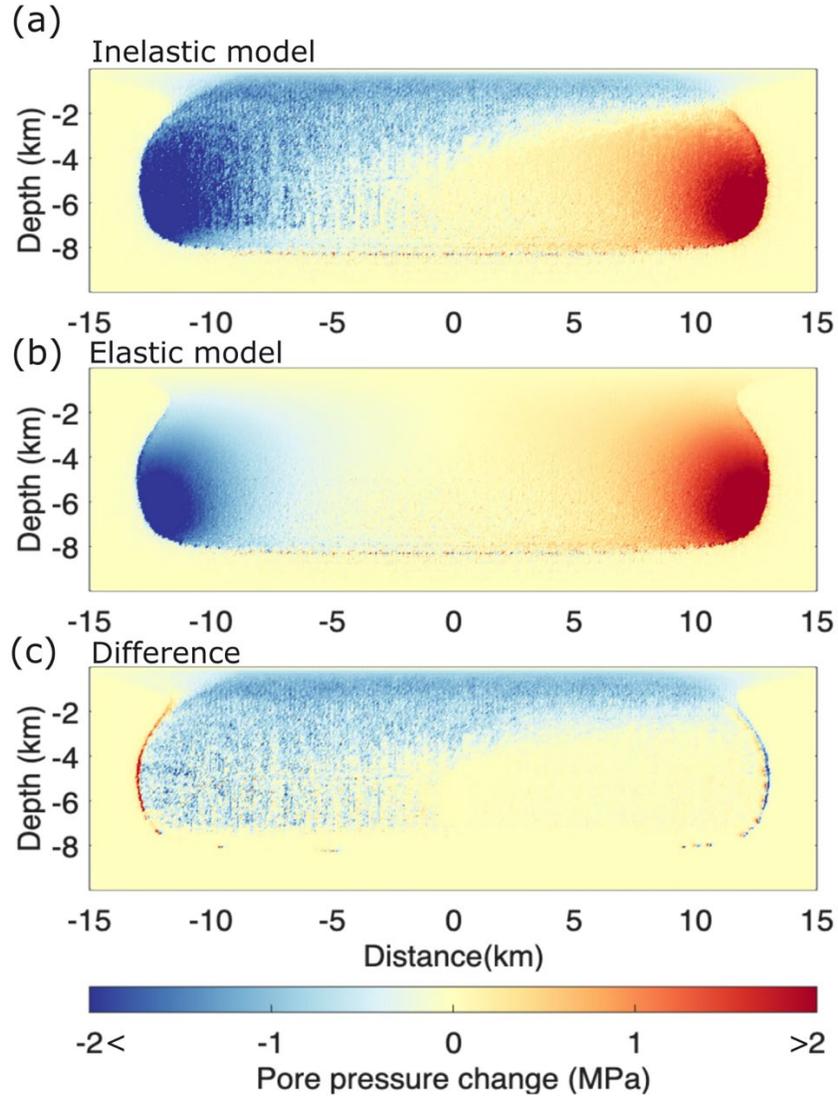

Figure 9. Coseismic undrained pore pressure change obtained from the earthquake dynamic rupture model on nearest slice (x=-30m) from the modeled fault plane (x=0 m) for (a) inelastic model, (b) elastic model, and (c) the pressure difference between (a) & (b). Negative x is in dilation quadrant (see Fig 2a).

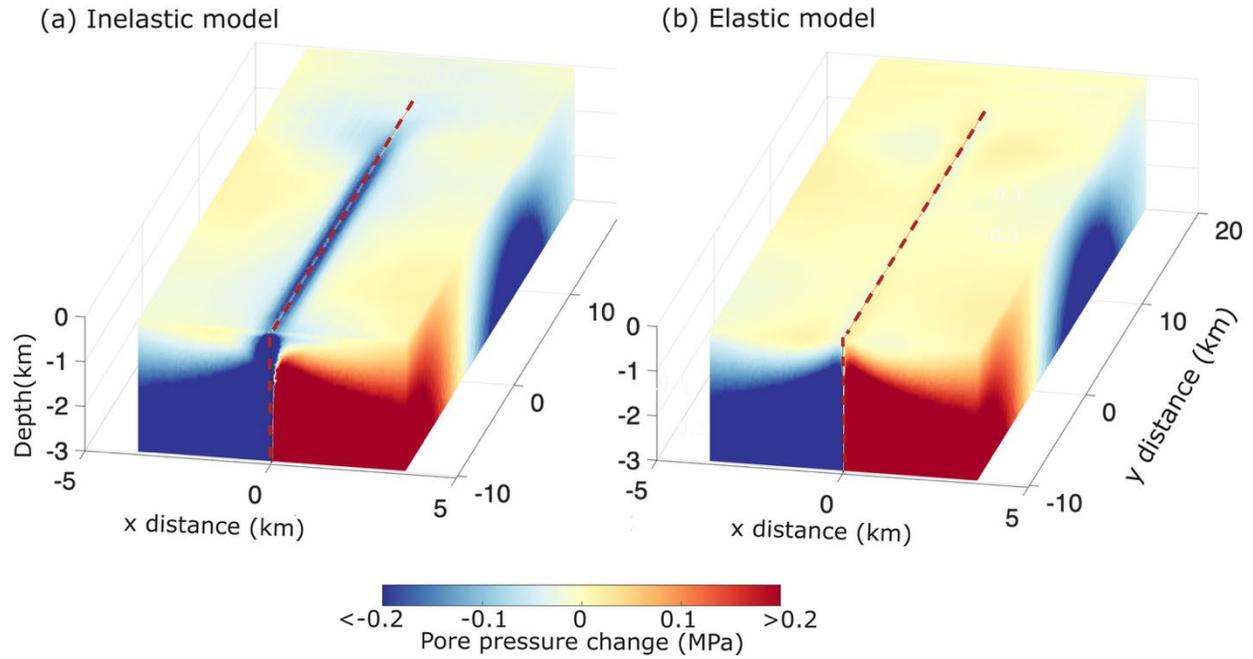

Figure 10. Undrained pore pressure changes in the coseismic fault-zone and off-fault regions. (a) results from the inelastic model, and (b) results from the elastic model. Dashed red line indicates the fault plane.

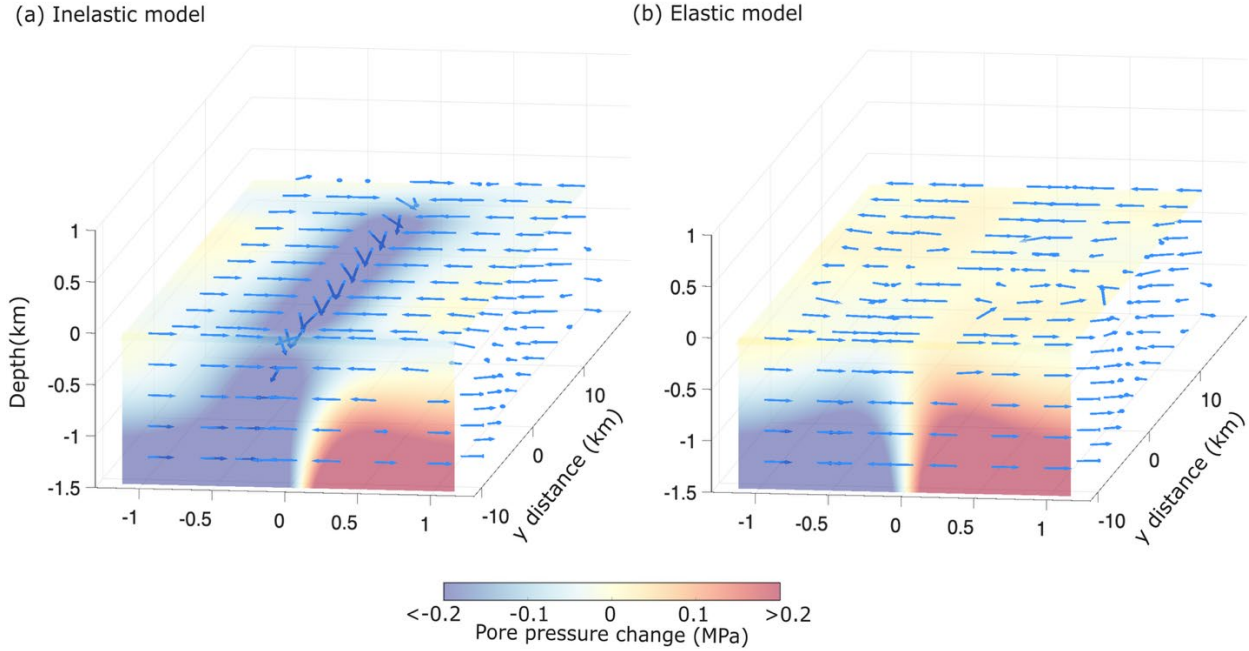

Figure 11. 3-D instantaneous groundwater flow induced by the fault rupture for (a) inelastic model and (b) elastic model. In figure (a), the coseismic pore pressure change and permeability enhancement allows inflow migration towards the fault zone whereas water flow is only controlled by the poroelastic pore pressure in figure (b). The arrows only point the flow direction but not showing the magnitudes.

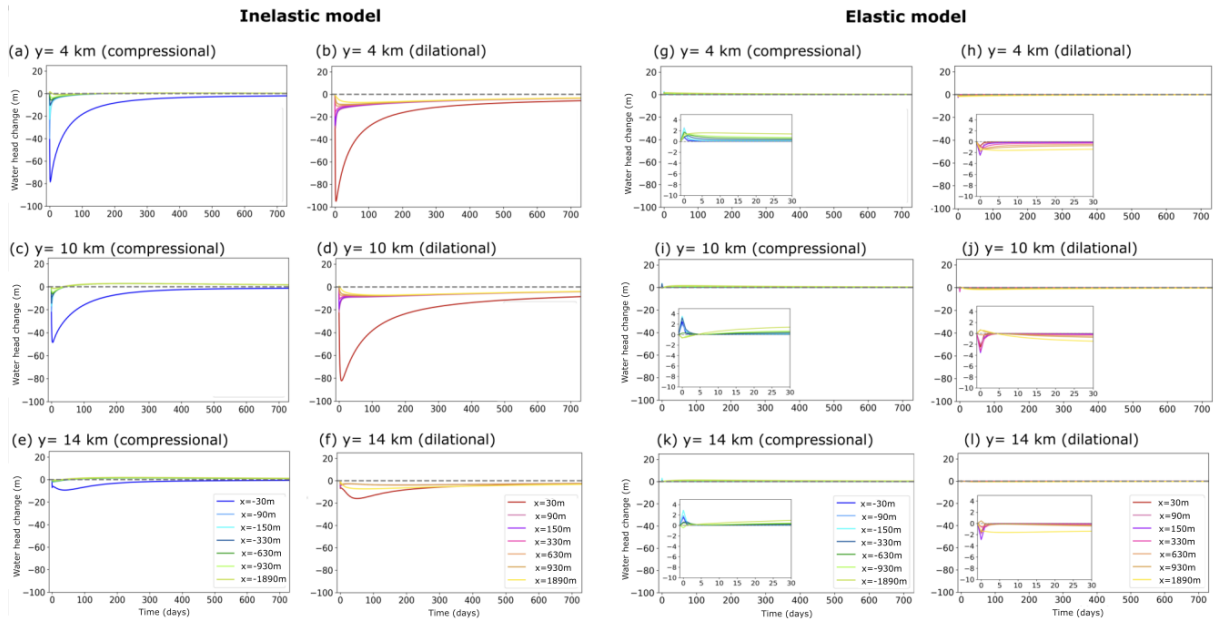

Figure 12. Time series of the modeled water levels measured along different fault-crossing profile (depth = 150 m) and compared between the two models. The left panels are water levels from the inelastic model and the right panels are from the elastic model. Black dashed line indicates water level change =0 (a) water levels at different fault-crossing distances (x values) in the fault-compressional quadrant at y=4 km. (b) similar to (a), but the water levels are sampled in the fault-dilational quadrant. Figure (c)-(f) are plotted in the same manner but for different y locations: (c) & (d) for y=10 km, (e) & (f) for y=14 km. Similarly, we plotted the water level responses for the elastic model. (g) & (h) water levels in different fault strain quadrants at y=4 km, (i) & (j) water levels in different fault strain quadrants at y=10 km, and (k) & (l) water levels in different fault strain quadrants at y=14 km. Inserted figures show the zoom-in plots focused on the early postseismic period between day 0-30.

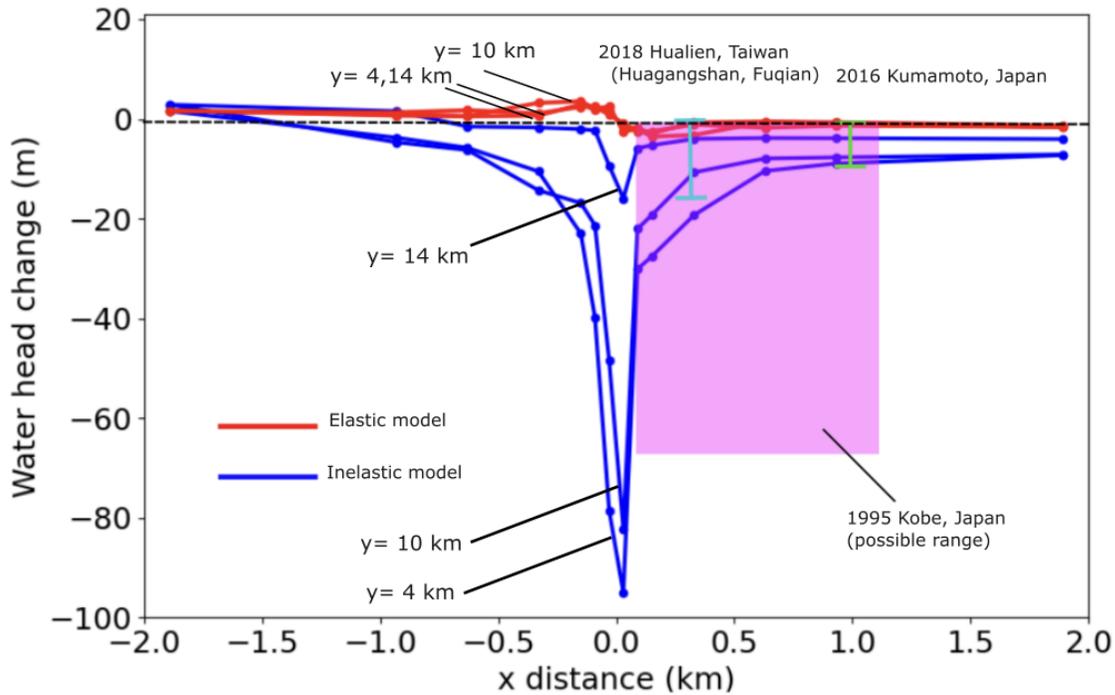

Figure 13. Simulated maximum water level response for inelastic (blue lines) and elastic (red lines) at different fault-crossing distance (x values) for two-year postseismic period. Some selected groundwater records (drawdowns) from strike-slip earthquakes are plotted for comparison (1995 Kobe, Japan: Tokunaga, 1999; 2016 Kumamoto, Japan: Hosono et al., 2019; 2018 Hualien, Taiwan: Hung et al., 2024; see Table 1 for details).